\documentclass[reprint,aps,twocolumn,superscriptaddress,10pt]{revtex4-1}
\usepackage{amsmath}
\usepackage[english]{babel}
\usepackage{graphicx}
\usepackage{color}
\usepackage[dvipsnames]{xcolor}
\usepackage{caption}
\definecolor{darkblue}{rgb}{0,0,0.8}

\usepackage{bbold}

\usepackage{soul}

\newcommand{\Moire}{Moir\'e}
\newcommand{\moire}{moir\'e}

\def \ETH{Institute for Quantum Electronics, ETH Z\"urich, CH-8093 Z\"urich, Switzerland}
\def \NIMSRCFM{Research Center for Functional Materials, National Institute for Materials Science, Tsukuba, Ibaraki 305-0044, Japan}
\def \NIMSICMN{International Center for Materials Nanoarchitectonics, National Institute for Materials Science, Tsukuba, Ibaraki 305-0044, Japan}

\def \MPI{Max Planck Institute of Quantum Optics, 85748 Garching, Germany\\ 
$^\dagger$These authors contributed equally to this work}
\def \MCQST{M\"unchen Center for Quantum Science and Technology, Schellingstrasse 4, 80799 M\"unich, Germany}
\def \TUM{Department of Physics and Institute for Advanced Study, Technical University of Munich, 85748 Garching, Germany}

\begin{document} 

\title{Optical signatures of charge order in a Mott-Wigner state}

\author{Yuya Shimazaki$^\dagger$}
\affiliation{\ETH}

\author{Clemens Kuhlenkamp$^\dagger$}
\affiliation{\ETH}
\affiliation{\TUM}
\affiliation{\MCQST}

\author{Ido Schwartz$^\dagger$}
\affiliation{\ETH}

\author{Tomasz Smolenski$^\dagger$}
\affiliation{\ETH}

\author{Kenji Watanabe}
\affiliation{\NIMSRCFM}

\author{Takashi Taniguchi}
\affiliation{\NIMSICMN}

\author{Martin Kroner}
\affiliation{\ETH}

\author{Richard Schmidt}
\affiliation{\MCQST}
\affiliation{\MPI}

\author{Michael Knap}
\affiliation{\TUM}
\affiliation{\MCQST}

\author{Ata\c{c} Imamo\u{g}lu}
\email{imamoglu@phys.ethz.ch}
\affiliation{\ETH}

\begin{abstract}
{The elementary optical excitations in  two dimensional semiconductors hosting itinerant electrons are attractive and repulsive polarons -- excitons that are dynamically screened by electrons. Exciton-polarons have hitherto been studied in translationally invariant degenerate Fermi systems.
Here, we show that electronic charge order breaks the excitonic translational invariance and leads to a direct optical signature in the exciton-polaron spectrum. Specifically, we demonstrate that new optical resonances appear due to spatially modulated interaction between excitons and electrons in an incompressible Mott state. Our observations demonstrate that resonant optical spectroscopy provides an invaluable tool for studying strongly correlated states, such as Wigner crystals and density waves, where exciton-electron interactions are modified by the emergence of new electronic charge or spin order.}
\end{abstract}

\maketitle

\begin{figure*}[ht!]
	\includegraphics[width=0.99\textwidth]{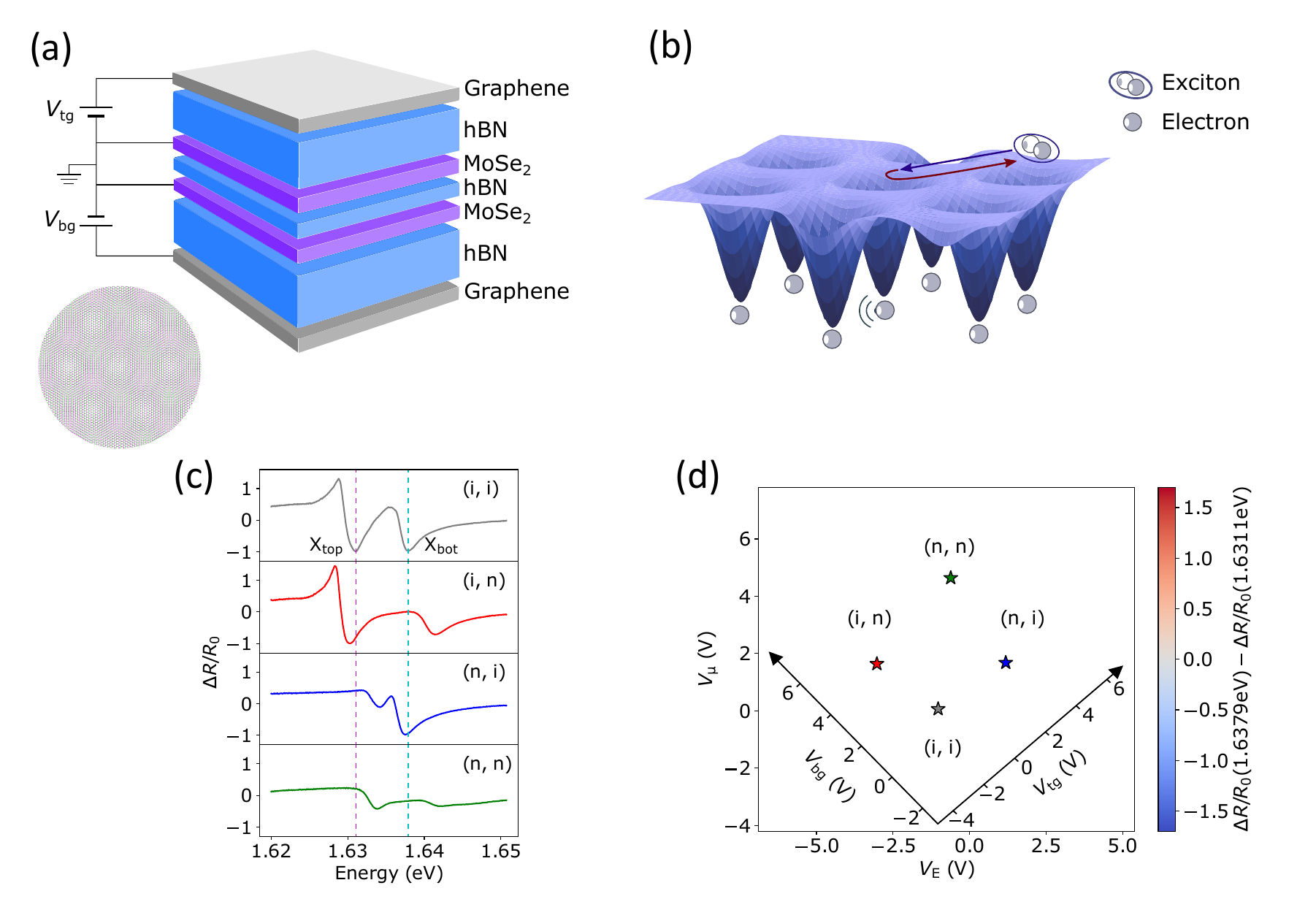}
	\caption{\Moire\ lattice in $\rm MoSe_2/hBN/MoSe_2$ and basic characterization.
	(a) Schematic picture of the device structure.
	The left bottom picture shows a \moire\ lattice.
	(b) Schematic picture of the potential for excitons created by electrons
	trapped in a \moire\ lattice.
	(c) Differential reflectance spectrum for each charge configuration.
	"i" and "n" indicate neutral and electron doped charge configurations,
	and shown in the order of (top, bottom).
	(d) Colormap of layer contrasted differential reflectance signal
	($\Delta R/R_0(1.6379~{\rm eV}) - \Delta R/R_0(1.6311~{\rm eV})$).}
	\label{fig:Fig1}
\end{figure*}

\section{Introduction} \label{sec:Sec1}

Bilayer graphene and transition metal dichalcogenide (TMD) heterostructures have emerged as fascinating new platforms to realize and probe exotic phases of quantum matter. Introducing a twist angle or lattice constant mismatch between the two layers allows for combining a \moire\ super-lattice and long-range Coulomb interactions, which in turn leads to the formation of strongly correlated electronic states~\cite{Cao2018a, Cao2018, Yankowitz2019}. Examples include the interplay between superconductivity and correlated insulators, as well as extended Mott-Wigner states, which exhibit translational symmetry breaking. In contrast to twisted bilayer graphene, TMD based heterostructures allow for the use of resonant optical reflection spectroscopy to study the signatures of new electronic phases in van der Waals heterostructures~\cite{Tang2020, Regan2020, Shimazaki2020, Wang2020}. While these discoveries have sparked the development of several novel methods to characterize such states~\cite{Xu2020}, most of the reported signatures were not directly sensitive to the spatial correlations of the quantum state of electrons.

Here, we show a striking, yet generic optical signature of the emergence of periodic ordering of electrons in a lattice structure. Due to strong exciton-electron interactions characteristic of TMD monolayers~\cite{Xu2014, Wang2018, Fey2020}, the formation of a Mott or Wigner state of electrons~\cite{Camjayi2008} creates a periodic potential for excitons, thereby modifying the exciton-polaron~\cite{Sidler2017, Efimkin2017} spectrum. Our experiments show that when either or both of the layers host an electronic Mott-state, a new Umklapp exciton-polaron peak appears in the resonant reflection spectrum. The energy shift of this new resonance is determined by the lattice constant of the emerging Mott state and by the strength of exciton-electron interaction.

In our experiments, we study a twisted MoSe$_2$/hBN/MoSe$_2$ homobilayer structure, which exhibits an incompressible single-layer Mott state for unity-filling of the underlying electronic \moire\ potential~\cite{Shimazaki2020}. 
In comparison to hetero-bilayer structures~\cite{Tang2020, Regan2020}, the presence of monolayer hBN in between the MoSe$_2$ monolayers leads to two new features: first, the on-site and possibly inter-site Coulomb repulsion energy exceeds the strength of the \moire\ potential, which is drastically weakened by the hBN monolayer. Second, the energy difference between the electronic states in the two layers is tunable, resulting in a robust layer pseudo-spin degree-of-freedom that can be controlled using an applied vertical electric field. 

Before proceeding, we note that recent photoluminescence (PL) experiments allowed for the characterization of a static excitonic \moire\ potential in twisted heterobilayers \cite{Seyler2019, Tran2019, Jin2019, Alexeev2019, Andersen2019}. This periodic potential is particularly strong for inter-layer excitons~\cite{Yu2017, Wu2018a, Ruiz-Tijerina2019}: in structures where two different transition metal dichalcogenide (TMD) monolayers are in direct contact, new inter- and intra-layer excitonic resonances arising either from localization at high symmetry stacking points or from Umklapp processes were observed. In stark contrast to these earlier works, the periodic static \moire\ potential experienced by the excitons in our sample is weak as compared to the exciton linewidth due to the hBN barrier layer: as a consequence, we observe only a single intra-layer exciton resonance in the absence of electron or hole doping.

\section{Basic characterization} \label{sec:Sec2}

Figure~\ref{fig:Fig1}(a) shows the sample we studied in our experiments: two MoSe$_2$ layers are isolated by monolayer hBN and are encapsulated between two thick hBN layers. Few-layer graphene sheets on the top and bottom of the device are used as transparent gate electrodes; we apply top and bottom gate voltages ($V_{\rm tg}$ and $V_{\rm bg}$, respectively) while keeping the MoSe$_2$ layers grounded.
A small twist angle between the top and bottom MoSe$_2$ layers results in the formation of a \moire\ superlattice, as shown schematically in the left bottom part of Fig.~\ref{fig:Fig1}(a). In our previous work~\cite{Shimazaki2020} based on the same device, we estimated the twist angle to be around 0.8 degree, which corresponds to the \moire\ lattice constant $a_{\rm M} \sim 25~{\rm nm}$ (see also Supplemental  Material S2). In the limit where each \moire\ unit cell of a single layer is singly occupied, the trapped electrons form a triangular lattice [Fig.~\ref{fig:Fig1}(b)].

\begin{figure*}[ht!]
	\includegraphics[width=1.0\textwidth]{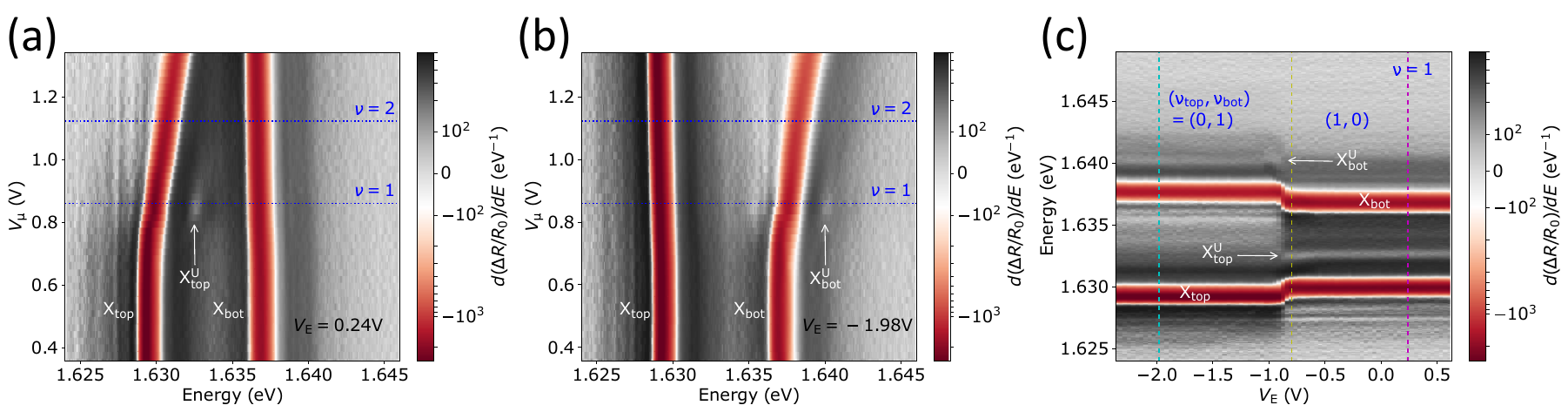}
	\caption{
	Umklapp exciton resonances at $\nu = 1$.
	(a), (b) $V_{\rm \mu}$ dependence of differential reflectance differentiated with respect to energy $E$ at $V_{\rm E} = 0.24~{\rm V}$ (a) and $V_{\rm E} = -1.98~{\rm V}$ (b).
	(c) $V_{\rm E}$ dependence of differential reflectance spectra at $\nu = 1$.
	The cyan, yellow, and magenta dashed lines correspond to $V_{\rm E} = -1.98 {\rm V}, -0.80~{\rm V}$ and $0.24~{\rm V}$, respectively.
	The scale of the color bars are logarithmic for $|d(\Delta R/R_0)/dE| > 10^2~{\rm eV^{-1}}$ and linear for $|d(\Delta R/R_0)/dE| < 10^2~{\rm eV^{-1}}$.
	}
	\label{fig:Fig2}
\end{figure*}

Figure~\ref{fig:Fig1}(c) shows the differential reflectance ($\Delta R/R_0 = (R - R_0)/R_0$) spectrum around the exciton resonances for  different charge configurations of the two MoSe$_2$ layers. Here, $R$ denotes the reflectance of the MoSe$_2$/hBN/MoSe$_2$ homobilayer structure, whereas $R_0$ is the background reflectance obtained from a region of the sample without MoSe$_2$ layers. All experiments depicted in the main text are performed using a cryogenic confocal microscope setup operated at a temperature $\sim$ 4~K. The results we present in Sections~\ref{sec:Sec2}~$\&~$\ref{sec:Sec3} are obtained in the absence of an external magnetic field ($B_z=0$). The gray curve in the upper panel depicts $\Delta R/R_0$ when both layers are in the charge-neutral regime. Accordingly, the two resonances correspond to the top and bottom layer excitons ($\rm X_{top}$ and $\rm X_{bot}$) whose energies differ by $\sim 8$~meV due to inhomogeneous strain~\cite{He2013, Conley2013, Zhu2013}.
When one of the layer is electron doped, exciton-electron interactions lead to a blue-shift of the repulsive polaron resonance frequency \cite{Sidler2017, Efimkin2017}. Since the  exciton-electron interactions are short-ranged, excitons in a given layer scatter predominantly with electrons in the same layer. 
By monitoring the magnitude of $\Delta R/R_0$ at a fixed energy close to either $\rm X_{top}$ or $\rm X_{bot}$ resonances, we can monitor the charge configuration of each layer. Since we do not consider the changes in the attractive polaron sector of the spectrum, we simply refer to repulsive polaron transitions as excitonic resonances in the following discussion. 

Figure~\ref{fig:Fig1}(d) shows the gate voltage dependence of the layer contrasted $\Delta R/R_0$ signal, revealing the charge configuration of the device~\cite{Shimazaki2020}. The color-coded plot shows the difference of $\Delta R/R_0$ signals at 1.6379~eV and 1.6311~eV which are indicated with cyan and magenta lines in Fig.~\ref{fig:Fig1}(c), respectively. We define the gate voltage axes as $V_{\rm E} = 0.5V_{\rm tg} - 0.5V_{\rm bg}$ and $V_{\rm \mu} = 0.4561V_{\rm tg} + 0.5439V_{\rm bg}$, which respectively correspond to changes of the top and bottom gate voltages that leave the chemical potential and the electric field invariant in the charge-neutral regime. The combinations of $V_{\rm E}$ and $V_{\rm \mu}$  used to obtain the spectra in Fig.~\ref{fig:Fig1}(c) are marked with color coded stars in Fig.~\ref{fig:Fig1}(d).

\begin{figure}[ht!]
	\includegraphics[width=0.5\textwidth]{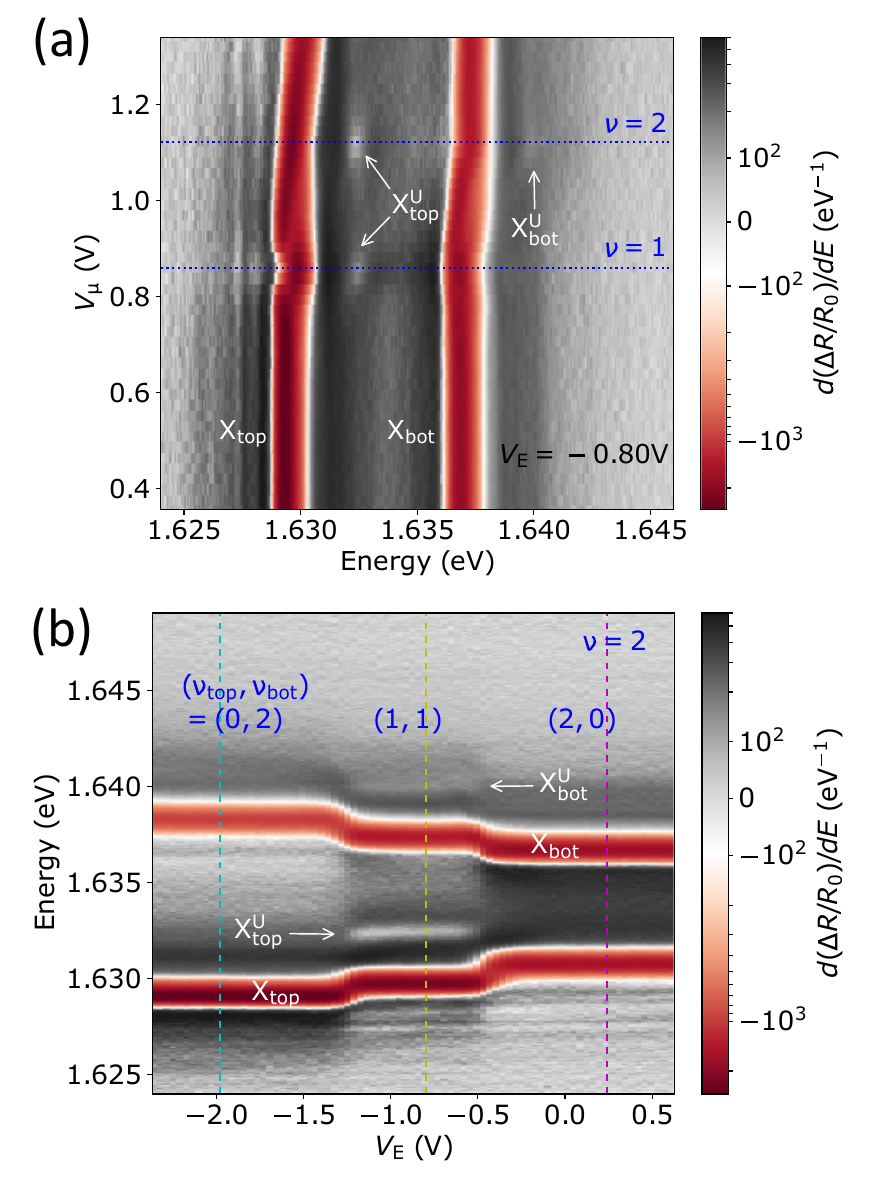}
	\caption{
	Umklapp exciton resonances at $\nu = 2$.
	(a) $V_{\rm \mu}$ dependence of differential reflectance differentiated with respect to energy $E$ at $V_{\rm E} = -0.80~{\rm V}$.
	(b) $V_{\rm E}$ dependence of differential reflectance spectra at $\nu = 2$.
	The cyan, yellow, and magenta dashed lines correspond to $V_{\rm E} = -1.98~{\rm V}, -0.80~{\rm V}$ and $0.24~{\rm V}$, respectively.
	The scale of the color bars are logarithmic for $|d(\Delta R/R_0)/dE| > 10^2~{\rm eV^{-1}}$ and linear for $|d(\Delta R/R_0)/dE| < 10^2~{\rm eV^{-1}}$.
	}
	\label{fig:Fig3}
\end{figure}

\section{Spectroscopic signature of exciton Umklapp scattering in Mott states} \label{sec:Sec3}

Strong electronic correlations become manifest in a half-filled \moire\ subband ($\nu = 1$), where each \moire\ site is occupied by a single electron. Since our structure has a layer degree of freedom, we specify the electron filling factor of top and bottom layers as $\nu_{\rm top}$ and $\nu_{\rm bot}$, with $\nu = \nu_{\rm top} + \nu_{\rm bot}$ (see Supplemental  Material S2 for the identification of filling).
In Fig.~\ref{fig:Fig2}(a), we show $V_{\rm \mu}$ dependence of reflectance spectrum at a fixed $V_{\rm E} = 0.24~{\rm V}$. To better visualize small signals, we take moving average of $\Delta R/R_0$ within a 0.24 meV energy window, and plot the derivative of $\Delta R/R_0$ with respect to energy.
The bottom layer remains neutral at this $V_{\rm E}$ and only the filling of the top layer is modulated along the $V_{\rm \mu}$ axis, as evidenced by the blue shift of $\rm X_{top}$ resonance.
Remarkably, we observe that once the top layer is doped to $\nu = 1$ where $(\nu_{\rm top}, \nu_{\rm bot}) = (1, 0)$,  an additional higher energy exciton resonance labeled as $\rm X_{top}^U$ appears. The estimated energy separation between $\rm X_{top}$ and $\rm X_{top}^{U}$, determined from resonant fluorescence data, is $\simeq 2.7$~meV (see Supplemental  Material S1).

The corresponding resonance also appears when the bottom layer is doped to $\nu=1$: Fig.~\ref{fig:Fig2}(b) shows the $V_{\rm \mu}$ dependence of $d(\Delta R/R_0)/dE$ at a fixed $V_{\rm E} = -1.98~{\rm V}$, where we only fill the bottom layer and the top layer remains charge neutral.
At $\nu = 1$ where $(\nu_{\rm top}, \nu_{\rm bot}) = (0, 1)$, we also observe a faint resonance labeled as $\rm X_{bot}^U$ on the blue side of $\rm X_{bot}$.
The energy separation between $\rm X_{bot}$ and $\rm X_{bot}^{U}$ is also estimated to be $\simeq 2.7$~meV.

Figure~\ref{fig:Fig2}(c) shows the $V_{\rm E}$ dependence of $d(\Delta R/R_0)/dE$ at $\nu = 1$. The sharp energy shift of the $\rm X_{top}$ and $\rm X_{bot}$ resonances around $V_{\rm E} \sim -0.9~{\rm V}$ signals an abrupt transfer of all electrons from one layer to another upon minute changes in the electric field $E_z$. As $\nu$ is increased or decreased away from $1$ (one electron per \moire\ site), the electrons are transferred from one layer to another gradually (see Fig. S3 in Supplemental  Material). These observations indicate the formation of an incompressible Mott insulator state at $\nu=1$~\cite{Shimazaki2020}: the abrupt transfer of electrons in turn indicates that the potential minima of the bottom and top layer \moire\ potentials for electrons are displaced from each other. By transferring the Mott insulating phase from the top ($(\nu_{\rm top}, \nu_{\rm bot}) = (1, 0)$) to the bottom ($(\nu_{\rm top}, \nu_{\rm bot}) = (0, 1)$) layer, we find the $\rm X_{top}^U$ resonance disappears and $\rm X_{bot}^U$ resonance appears. In contrast to strong $V_{\rm \mu}$ dependence, the $\rm X_{top}^U$ and $\rm X_{bot}^U$ resonances consistently exist for a wide range of $V_{\rm E}$ at $\nu = 1$.

Since the new excitonic resonance in the top (bottom) layer appears exclusively when $\nu_{\rm top} = 1$ ($\nu_{\rm bot} = 1$), we explain its origin as the emergence of a periodic potential for excitons generated by electrons in the singly-occupied Mott state. Exciton-electron interactions ensure scattering of resonantly generated excitons off this Mott-state-potential and lead to the emergence of excitonic bands in the new reduced exciton Brillouin zone. 
In the limit of weak exciton-electron scattering, the new excitonic resonances can be understood as Umklapp processes. In this limit the energy of the Umklapp scattered exciton is determined by the reciprocal lattice vector of the electron-induced potential alone, which is fixed by the \moire -periodicity. The estimated lattice constant of $\simeq 25$~nm for this sample~\cite{Shimazaki2020} yields a splitting of $\simeq 2.5$~meV, which is in very good agreement with the experimentally observed splitting of the Umklapp resonance (see Appendix B).

Before presenting a theoretical model for these experimental observations, we investigate the optical response when there are two electrons per \moire\ site ($\nu = 2$). 
Figure~\ref{fig:Fig3}(a) shows $V_{\rm \mu}$ dependence of $d(\Delta R/R_0)/dE$ at a fixed $V_{\rm E} = -0.80~{\rm V}$.
At this $V_{\rm E}$, indicated with a yellow dashed line in Fig.~\ref{fig:Fig2}(c), the system is filled as $(\nu_{\rm top}, \nu_{\rm bot}) = (1, 0)$ at $\nu = 1$, and $(\nu_{\rm top}, \nu_{\rm bot}) = (1, 1)$ at $\nu = 2$.
Consistent with the measurements depicted in Fig.~\ref{fig:Fig2}(a), we find the emergence of $\rm X_{top}^U$ at $\nu = 1$ in Fig.~\ref{fig:Fig3}(a).
Remarkably, at $\nu = 2$ where $(\nu_{\rm top}, \nu_{\rm bot}) = (1, 1)$, the Umklapp exciton resonances $\rm X_{top}^U$ and $\rm X_{bot}^U$ emerge simultaneously. This observation at $(\nu_{\rm top}, \nu_{\rm bot}) = (1, 1)$, unequivocally shows the essential role played by the Mott state of electrons in effecting a periodic potential for excitons.
The estimated energy separation between $\rm X_{top}$ and $\rm X_{top}^{U}$ is $\simeq 2.8$~meV; similarly, the separation between $\rm X_{bot}$ and $\rm X_{bot}^{U}$ is $\simeq 2.6$~meV. The magnitude of these energy separations are very  similar to the ones obtained at $\nu = 1$, indicating that the top and bottom layers have the same triangular electron lattices and that the periodicity is not affected by the simultaneous presence of Mott states in the two layers.
The similarity of energy separation for $\nu = 1$ and $\nu = 2$ also suggests that these new resonances are unlikely to originate from shake-up processes, since the excitation spectrum of a Mott state is expected to be sensitive to the filling.

Finally, we show the $V_{\rm E}$ dependence of $d(\Delta R/R_0)/dE$ at $\nu = 2$ in Fig.~\ref{fig:Fig3}(b). The yellow dashed line indicates $V_{\rm E}=-0.80~{\rm V}$; the plateau structure of the excitonic resonance energies for $-1.2 \le V_{\rm E} \le -0.60~{\rm V}$ shows the resilience of the layer occupancy to the applied electric field, indicating the formation of Mott insulating states simultaneously in both layers at $(\nu_{\rm top}, \nu_{\rm bot}) = (1, 1)$. Consistently, we observe top and bottom layer Umklapp excitonic resonances throughout the range of $V_{\rm E}$ for which  $(\nu_{\rm top}, \nu_{\rm bot}) = (1, 1)$. We also remark that the Umklapp resonances are absent for $(\nu_{\rm top}, \nu_{\rm bot}) = (2, 0),\ (0, 2)$ (which we also confirm from Fig.~\ref{fig:Fig2}(a), (b)), indicating that the electrons do not form a Mott state with 2 electrons per \moire\ potential site.

\begin{figure*}[ht!]
    \begin{centering}
\includegraphics[width=1\textwidth]{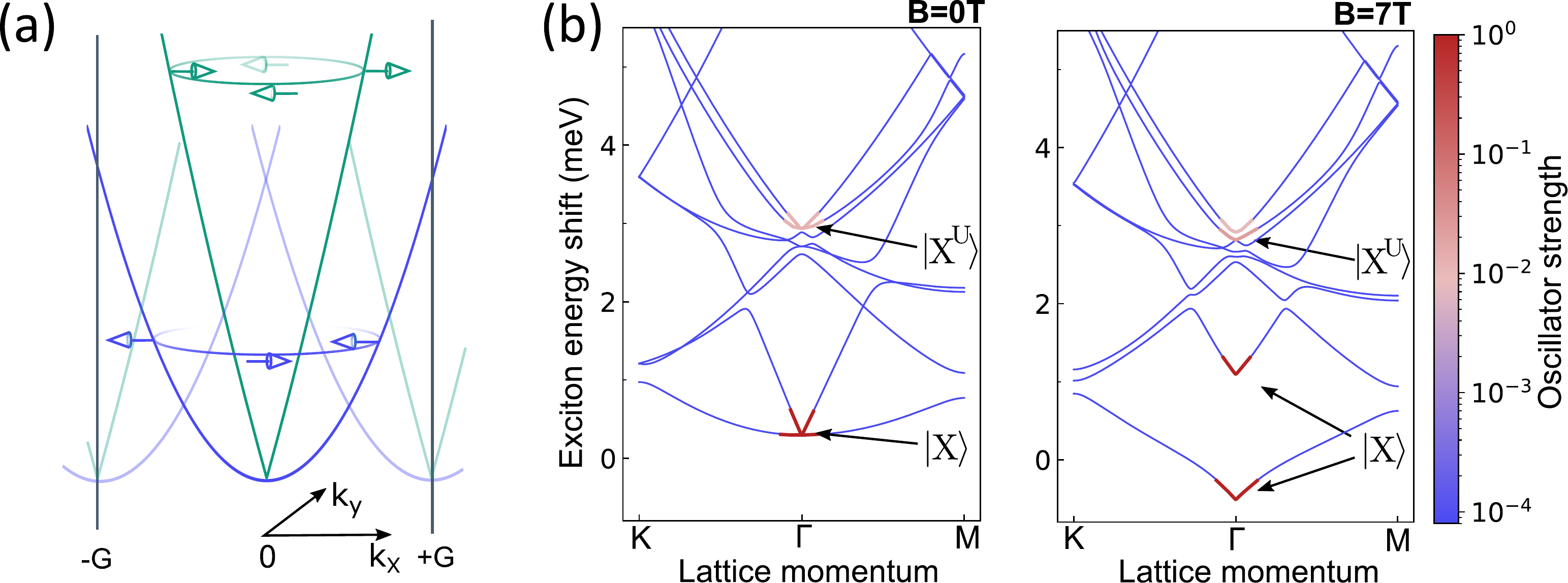}
	\end{centering}
	\caption{Exciton band structure. Spectrum of the effective band Hamiltonian of Eq.~\ref{Eq:hamiltonian_effective} for parameters given in the main text and $J=300$~meV. 
(a) Bare dispersion of mobile excitons. Strong inter-valley electron-hole exchange coupling splits the linearly polarized exciton modes and yields two branches with linear (green line) and parabolic (blue line) dispersion. For simplicity, we do not show that the degeneracy of the excitonic branches extend throughout the light-cone. The exciton valley pseudo-spin is shown as blue (green) arrows for the parabolic (linear) dispersion. Periodic potentials allow states connected by reciprocal lattice vectors to mix, yielding higher bands (dashed lines), as shown in an extended zone scheme. Higher bands of the linearly dispersing mode decouple due to their large energy detuning.
(b) Exciton bands along a path in the Brillouin zone. The oscillator strength of each state is indicated by the color bar, which is saturated for all blue lines. As photons carry almost vanishing momentum, only modes close to the $\Gamma$-point may obtain finite oscillator strength (we artificially extended the momentum range of the bright states for better visibility). While most states remain dark, only a single Umklapp band per polarization obtains sizable oscillator strength. (Left) Dispersion in the absence of magnetic field. We find an energy splitting between the Umklapp state and the main resonance of $3$~meV. This additional bright state carries 1.2\% of the oscillator strength of the bare exciton. The oscillator strength of higher bands is significantly suppressed due to the larger energy splitting, which reduces the coupling to light. (Right) Exciton dispersion for $B_z=7~{\rm T}$. While the main resonance splits significantly, the Umklapp peaks are only marginally affected by the magnetic field.}
	\label{fig:Fig4}
\end{figure*}

\section{Theoretical model} \label{sec:Sec4}

The appearance of additional higher-energy excitonic resonances in the optical spectrum can be described by a simple model of excitons moving in a periodic potential. For simplicity we are focusing only on a single layer which hosts the strongly correlated electronic state. As excitons are strongly bound, we consider them as mobile rigid quantum impurities. The two valleys of the TMD imprint as a (valley) pesudo-spin degree of freedom for the excitons. The optically active $\mathbf{k}=0$ excitons are $\sigma^+$ ($K$-valley) and $\sigma^-$ ($K'$-valley) polarized and are degenerate. However, for exciton momenta outside the light cone, the long-range electron-hole exchange interaction strongly couples the $K$ and $K'$ valley excitons \cite{Yu2014, Glazov2014}. In the pseudo-spin basis, the excitons are described by the Hamiltonian:
\begin{equation}
H_0 = \sum_{\boldsymbol{k}}  \frac{\hbar^2 \boldsymbol{k}^2}{2m_{\rm X}} + \frac{|\boldsymbol{k}|}{|K|}J\begin{pmatrix}
0 & e^{-i2\theta} \\
e^{+i2\theta} & 0 
\end{pmatrix} +\frac{|\boldsymbol{k}|}{|K|}J,
\label{Eq:dispersion_excitons}
\end{equation}
where $\hbar$ is Planck's constant, $\theta = \mathrm{atan} (k_y/k_x)$ and $J\sim 1$~eV \cite{Qiu2015}. The second and the third terms of Eq.~\ref{Eq:dispersion_excitons} describe inter-valley and intra-valley exchange interaction, respectively. We remark that the exchange coupling is not easily accessible experimentally and its exact value is therefore uncertain. We expect, however, that the experimentally relevant coupling is likely to be reduced by dielectric screening due to the hBN encapsulation, as well as due to screening by electrons. We have confirmed that our conclusions are unaffected, even if we assume a relatively low value of $J\sim 150$~meV. The exciton dispersions are given by $E(\boldsymbol{k}) = \frac{\hbar^2\boldsymbol{k}^2}{2m_{\rm X}} +  \frac{|\boldsymbol{k}|}{|K|}J \pm  \frac{|\boldsymbol{k}|}{|K|}J $ and are shown in  Fig.~\ref{fig:Fig4}(a). Exchange interactions split the polarizations into two branches with parabolic and linear dispersion. The linearly dispersing excitons fall on a steep cone, which leads to a large energy detuning from the parabolic branch.

The interaction between electrons and excitons is modelled by an effective repulsive contact interaction, which is justified, as we limit our discussion to features that appear at low energies $0 \le E \le 5$~meV. The electron-exciton interaction Hamiltonian then takes the form:
\begin{equation}
\begin{aligned}
H^{\mathrm{int}}_{e-x} &=\int d^2r\;\lambda^{e-x} \left[\hat{n}_e(\boldsymbol{r}) \hat{n}^+_{\rm X}(\boldsymbol{r}) +  \hat{n}_e(\boldsymbol{r})\hat{n}^-_{\rm X}(\boldsymbol{r})\right],
\end{aligned}
\label{Eq:hamiltonian1}
\end{equation}
where $\hat{n}^{\pm}_{\rm X}$ is the density operator of excitons with pseudo-spin $\sigma=\pm$ and $\hat{n}_e$ is the density operator of the electrons. We assume that the strength $\lambda^{e-x}$ of the interaction is the same for both polarizations and that the interaction itself is short-range and repulsive. The delocalized nature of optically generated excitons ensures that the electronic Mott-Wigner state, appearing due to an interplay between long-range Coulomb interactions and the moir\'e potential, is largely unperturbed by the exciton-electron interactions: this is verified as the experimentally observed signatures of incompressible states are independent of the white-light intensity. To simplify the analysis, we assume that electrons are frozen in the Mott-Wigner ground state and describe the excitonic degrees of freedom alone by an effective Hamiltonian
\begin{equation}
\begin{aligned}
H_{\mathrm{eff}} =& \sum_{\boldsymbol{k}}  \frac{\hbar^2\boldsymbol{k}^2}{2m_{\rm X}} + \frac{|\boldsymbol{k}|}{|K|}J\begin{pmatrix}
0 & e^{-i2\theta} \\
e^{+i2\theta} & 0 
\end{pmatrix} +\frac{|\boldsymbol{k}|}{|K|}J \\
&+ V(\boldsymbol{r})  \;
\mathbb{1}_{2\times 2},
\end{aligned}
\label{Eq:hamiltonian_effective}
\end{equation}
where $V(\boldsymbol{r})  = \lambda^{e-x}\langle \hat{ n}_e(\boldsymbol{r})\rangle$ is the potential, $\langle \hat{ n}_e(\boldsymbol{r})\rangle$ is the expectation value of the electron density and $m_{\rm X}$ is the exciton mass.
For the exciton mass we use $m_{\rm X} = m_e^* + m_h^* = 1.3\,m_e$, where $m_e^* = 0.7\,m_e$~\cite{Larentis2018} and $m_h^* = 0.6\,m_e$~\cite{Zhang2014} are the effective masses of electrons and holes  ($m_e$ is the free electron mass).
By using $H_{\mathrm{eff}}$, we neglect the dynamical dressing of excitons by virtual excitations out of the Mott state and take the effect of electronic correlations into account as a spatially dependent Hartree-shift of the exciton energy, proportional to the expectation value of the electron density $\langle \hat{ n}_e(\boldsymbol{r})\rangle$. While the precise shape of the potential seen by the excitons is determined by the electronic density and the moir\'e potential, we phenomenologically fix the Hartree shift induced by $\lambda^{e-x}$ to match the experimentally measured blue-shift of the repulsive-polaron resonance induced by the electrons at low densities. We thereby determine $\lambda^{e-x} = 2.1 \times 10^{-12}~{\rm meV \cdot cm^{2}}$ to produce an excitonic blue-shift of $0.4$~meV at fillings slightly away from $\nu=1$, where the electron density is expected to be homogeneous.

We solve the effective single-particle model of Eq.~\ref{Eq:hamiltonian_effective} assuming a gaussian density profile for electrons
\begin{equation}
 \langle \hat{n}_e(\boldsymbol{r})\rangle =  \frac{1}{2 \pi \xi^2} \sum_{\boldsymbol{R}} e^{- \frac{1}{2 \xi^2} \left(\boldsymbol{r} - \boldsymbol{R}\right)^2} ,  
\end{equation}
where $\boldsymbol{R}$ is a triangular lattice vector and $\xi$ characterizes the extent of the electronic wave functions around the \moire\ sites \cite{Yoshioka1979}. We show the band-structure resulting from the electron-induced potential in Fig.~\ref{fig:Fig4}(b) for an electronic lattice with a lattice constant of $25$~nm and $\xi=$ 4~nm.  While the precise localization of the electrons is not known, we have checked that our results remain in reasonable agreement with the experiment for $\xi$'s within 3~nm $< \xi < $ 6~nm. The new bands in the \moire\ Brillouin zone appear as a consequence of the periodic excitonic potential, as momentum is no longer a good quantum number and is conserved only up to reciprocal lattice vectors $\mathbf{G}_m$. In particular this implies that excitons carrying a reciprocal lattice momentum $\boldsymbol{k}=\mathbf{G}$ now mix with the optically active $\boldsymbol{k}=0$ excitons via Umklapp scattering. 

The oscillator-strength of the Umklapp states is given by their zero-momentum exciton content. We numerically determine the relative oscillator strength of the first optically active Umklapp band and find its relative oscillator strength to be:
\begin{equation}
 \frac{|\langle \boldsymbol{k}=0|\mathrm{X}^{\rm U}(\boldsymbol{\Gamma})\rangle|^2}{ |\langle  \boldsymbol{k}=0|\mathrm{X}(\boldsymbol{\Gamma}) \rangle|^2}\simeq 1.2\%, \\
 \label{Eq:osc_strength}
\end{equation}
where $|\mathrm {X}\rangle$ and $|\mathrm{X^U}\rangle$ are the exciton and the bright Umklapp-exciton states of a given polarization, while $\boldsymbol{\Gamma}$ labels vanishing lattice momentum. In Eq.~\ref{Eq:osc_strength}, the vector $|\boldsymbol{k}\rangle$ is a plane wave state of the exciton with proper momentum $\boldsymbol{k}$. The amount of mixing and the band splittings are determined by the strength of the potential and the density profile of the electrons. However, the $\mathrm{C_6}$ symmetry of the triangular potential restricts mixing with the $\boldsymbol{k}=0$ modes and hence the number of possible bright states: only states which  transform identically under rotations (and are hence grouped in the same $\mathrm{C}_6$ representation) can have non-vanishing matrix elements. In our case we find only two states among the first Umklapp band with $\mathrm{C_6}$ eigenvalues ${\rm exp}(il\pi/3)$ where $l=\pm 1$, which are circularly polarized and mix with the $|\boldsymbol{k}=0 \rangle$ excitons to become bright. The oscillator strength of the bright resonances decreases rapidly as the Umklapp energy increases; this feature renders only the first Umklapp band effectively observable in experiments.

Both, the appearance of a single Umklapp line per polarization and its estimated oscillator strength, are in good agreement with our experimental observations and confirm that Umklapp exciton/repulsive-polaron resonances provide a direct probe of the periodic structure of the strongly correlated Mott-Wigner state.

\section{Magnetic field dependence of Umklapp states} \label{sec:Sec5}

\begin{figure*}[ht!]
	\includegraphics[width=0.99\textwidth]{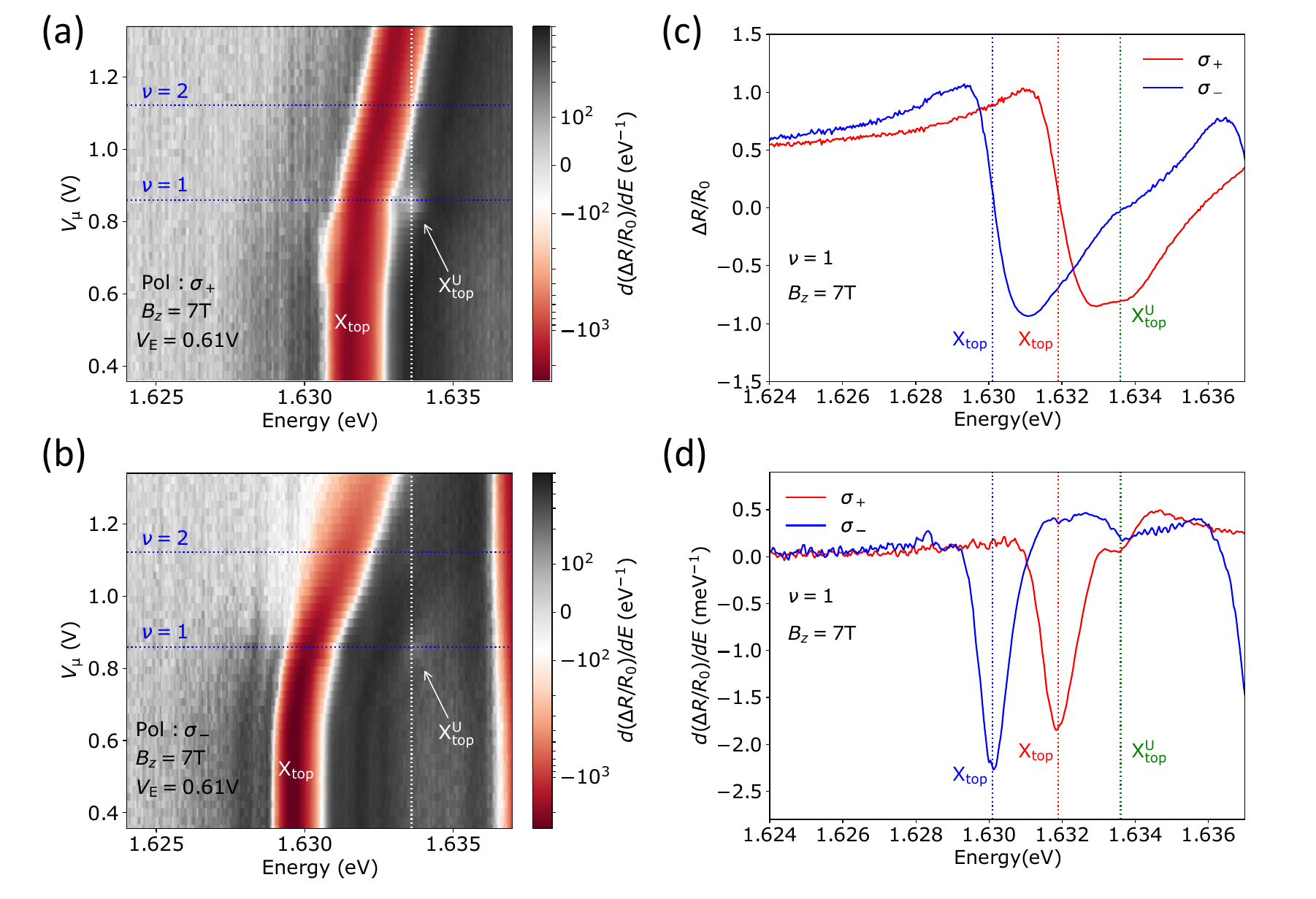}
	\caption{
	Exciton and Umklapp exciton resonances at $\nu = 1$ under magnetic field $B_z = 7~{\rm T}$.
	(a), (b) $V_{\rm \mu}$ dependence of differential reflectance differentiated with respect to energy $E$ in $\sigma_+$ (a) and $\sigma_-$ (b) polarization.
	(c), (d) Line cuts of the reflectance spectrum (c) and its energy differentiation (d) at $\nu = 1$, $V_{\rm E} = 0.61~{\rm V}$.
	}
	\label{fig:Fig5}
\end{figure*}

We now extend our analysis to finite magnetic fields, $B_z$. As excitons are strongly bound and charge-neutral, they only couple via an effective Zeeman term with a g-factor of  $\sim 4$~\cite{Li2014, Srivastava2015, Aivazian2015, MacNeill2015}. We neglect the momentum dependence of the Zeeman shift and introduce it by adding a (momentum-independent) term $\hat{H}_{\mathrm{Z}}=2 \mu_{\rm B} B_z \sigma_z$ to the Hamiltonian of Eq.~\ref{Eq:hamiltonian_effective}; here $\mu_{\rm B}$ is the Bohr magneton. Since electron-hole exchange interaction leads to a large energy splitting of linearly polarized longitudinal and trasverse exciton branches for exciton momenta~$\sim$~reciprocal lattice momenta, Umklapp states react only weakly to magnetic fields perpendicular to the sample. This allows for a unique and unambiguous test for the finite momentum character of the observed umklapp resonances: while the main peaks will acquire a splitting of $\sim 4\mu_{\rm B} B_z$, the Umklapp peaks will exhibit a much smaller splitting. To illustrate this, we show the full band structure of the exciton for a magnetic field of $7~$T in Fig.~\ref{fig:Fig4}(b) (right).

We experimentally confirmed the suppression of Zeeman splitting for Umklapp states for $B_z$ = 7~T. We performed polarization resolved $ \Delta R/R_0$ measurements and observe the emergence of Umklapp states ($\rm X_{top}^U$) in both $\sigma_+$ [Fig. \ref{fig:Fig5}(a)] and $\sigma_-$ [Fig. \ref{fig:Fig5}(b)] polarization around $\nu = 1$ where $(\nu_{\rm top}, \nu_{\rm bot}) = (1, 0)$. Fig. \ref{fig:Fig5}(c) and \ref{fig:Fig5}(d) show the line cuts of the $ \Delta R/R_0$ spectrum and its derivative with respect to emission energy at $\nu = 1$ where $(\nu_{\rm top}, \nu_{\rm bot}) = (1, 0)$.
Compared to the large Zeeman splitting of $\rm X_{top}$ of $\simeq 2$~meV, the energy splitting of $\rm X_{top}^U$ is much smaller than $1$~meV. On the one hand, the observation of vanishingly small Zeeman splitting of the high energy resonances appearing at $\nu = 1$ confirm their identification as Umklapp peaks. On the other hand, our measurements provide a direct evidence for the important role played by long-range electron-hole exchange interaction for high momentum exciton states.

\section{Discussions} \label{sec:Sec6}

In the theoretical analysis we treated the electronic lattice as rigid, and neglected dynamical screening effects such as the distortion of the electronic lattice by the impurity through collective excitations of the Mott-Wigner state. Nevertheless,  we expect our static potential model to capture the essential physical processes provided that the first Umklapp bands appear below the excitation gap of the electronic state. We estimate however, that this condition is only marginally satisfied in our experiments and a quantitative agreement with experiments would require the development of a theory of mobile quantum impurities in a strongly interacting Fermi system.

We also remark that our calculations assumed a perfect electronic lattice. Strictly speaking, this assumption is not justified since we find the electrons in the Mott-Wigner state to be spin-unpolarized and the electron-exciton scattering is known to be spin-dependent. Consequently, different exciton-electron scattering channels give rise to similar, but not identical, repulsive interactions~\cite{Fey2020}. We nevertheless expect our calculation of the Umklapp resonances to be qualitatively accurate despite the apparently random electron spin orientation in the \moire\ lattice sites, since it is known from the theory of alloys that band structures could remain robust, and without sizeable broadening of the electronic states in the $\Gamma$-valley, even in the presence of disorder~\cite{Popescu2010}.

Our observation that the $\sigma^+$ and $\sigma^-$ umklapp excitons remain nearly degenerate at $B_z=7$~T  is a direct consequence of strong inter-valley exchange coupling, which ensures that finite momentum excitons are in-plane linearly polarized, with a longitudinal-transverse energy splitting that is approximately an order of magnitude larger than the Zeeman energy of $\sim 2$~meV. This observation unequivocally demonstrates that the exciton resonances that appear at $\nu = 1$ originate from finite-momentum excitons and concurrently allow us to access the physics of high momentum excitons.

We also remark that there are new repulsive polaron resonances and a fine structure of the attractive polaron resonance for filling factors $\nu \geq 3$ (see Supplemental  Material S4). While we currently cannot provide an explanation for these features, we emphasize that even the underlying electronic state for these high electron densities is so far not understood. 

We emphasize that previously reported optical spectroscopy of Mott-Wigner states revealed signatures of incompressibility, but a direct evidence for the presence of charge order has been elusive. The observation of the Umklapp exciton resonance that we report on the other hand, is a direct consequence of the emergence of charge-density-wave order. In particular, we expect the Umklapp peak to appear even for compressible states with charge order and can provide a direct evidence for spontaneous breaking of translational invariance due to formation of a Wigner crystal~\cite{Wigner1934, Grimes1979, Cooper1996, Knorzer2020}, charge density waves or bubble phases~\cite{Koulakov1996, Moessner1996} in a defect-free monolayer without an external potential. The different interaction strength between excitons and electrons in different valleys in turn could be the basis for optically detecting spin order.

\section*{Acknowledgements}
This work was supported by the Swiss National Science Foundation (SNSF) under
Grant No. 200021-178909/1 and the European Research Council (ERC) Advanced Investigator Grant
(POLTDES). Y.S. acknowledges support from the Japan Society for the Promotion of Science (JSPS) overseas research fellowships. K.W. and T.T. acknowledge support from the Elemental Strategy Initiative conducted by MEXT, Japan, A3 Foresight by JSPS and CREST (grant no. JPMJCR15F3) and JST. We furthermore acknowledge support from the Technical University of Munich - Institute for Advanced Study, funded by the German Excellence Initiative, the European Union FP7 under grant agreement 291763, the Deutsche Forschungsgemeinschaft (DFG, German Research Foundation) under Germany's Excellence Strategy--EXC-2111--390814868, the European Union’s Horizon 2020 research and innovation programme (grant agreement No. 85116), and from the DFG grants No. KN1254/1-1 and No. KN1254/1-2.

\section*{Appendix A: Reflectance measurement}
We performed measurements with a cryogenic confocal microscope setup at a temperature $\sim$ 4~K. For the reflectance measurements, we illuminated the sample with a single-mode fibre-coupled broadband light-emitting diode with a centre wavelength of 760~nm and a bandwidth of 20~nm.

\section*{Appendix B: Umklapp exciton energy for weak exciton-electron interaction limit}
The energy separation between the exciton and the first Umklapp scattered exciton in the weak limit of exciton-electron interaction is given by the following expression,
\begin{equation}
    \Delta E_{\rm X^U-X} = \frac{\hbar^2|{\mathbf G_1}|^2}{2m_{\rm X}},
\end{equation}
where $\mathbf G_1$ is the lattice vector in Fig.~\ref{fig:Fig4}(c) which is $|{\mathbf G_1}| = 4\pi/(\sqrt{3}a_{\rm M})$, $a_{\rm M}$ is the \moire\ periodicity, and $m_{\rm X}$ is the exciton mass.
We use electron effective mass $m_e^* = 0.7m_e$~\cite{Larentis2018} and hole effective mass $m_h^* = 0.6m_e$~\cite{Zhang2014}, which gives $m_{\rm X} = m_e^* + m_h^* = 1.3m_e$, with $m_e$ denoting the free electron mass. For $a_{\rm M} = 25~{\rm nm}$, we obtain $\Delta E_{\rm X^U-X} = 2.5~{\rm meV}$.

\clearpage


\begin{thebibliography}{42}%
\makeatletter
\providecommand \@ifxundefined [1]{%
 \@ifx{#1\undefined}
}%
\providecommand \@ifnum [1]{%
 \ifnum #1\expandafter \@firstoftwo
 \else \expandafter \@secondoftwo
 \fi
}%
\providecommand \@ifx [1]{%
 \ifx #1\expandafter \@firstoftwo
 \else \expandafter \@secondoftwo
 \fi
}%
\providecommand \natexlab [1]{#1}%
\providecommand \enquote  [1]{``#1''}%
\providecommand \bibnamefont  [1]{#1}%
\providecommand \bibfnamefont [1]{#1}%
\providecommand \citenamefont [1]{#1}%
\providecommand \href@noop [0]{\@secondoftwo}%
\providecommand \href [0]{\begingroup \@sanitize@url \@href}%
\providecommand \@href[1]{\@@startlink{#1}\@@href}%
\providecommand \@@href[1]{\endgroup#1\@@endlink}%
\providecommand \@sanitize@url [0]{\catcode `\\12\catcode `\$12\catcode
  `\&12\catcode `\#12\catcode `\^12\catcode `\_12\catcode `\%12\relax}%
\providecommand \@@startlink[1]{}%
\providecommand \@@endlink[0]{}%
\providecommand \url  [0]{\begingroup\@sanitize@url \@url }%
\providecommand \@url [1]{\endgroup\@href {#1}{\urlprefix }}%
\providecommand \urlprefix  [0]{URL }%
\providecommand \Eprint [0]{\href }%
\providecommand \doibase [0]{http://dx.doi.org/}%
\providecommand \selectlanguage [0]{\@gobble}%
\providecommand \bibinfo  [0]{\@secondoftwo}%
\providecommand \bibfield  [0]{\@secondoftwo}%
\providecommand \translation [1]{[#1]}%
\providecommand \BibitemOpen [0]{}%
\providecommand \bibitemStop [0]{}%
\providecommand \bibitemNoStop [0]{.\EOS\space}%
\providecommand \EOS [0]{\spacefactor3000\relax}%
\providecommand \BibitemShut  [1]{\csname bibitem#1\endcsname}%
\let\auto@bib@innerbib\@empty
\bibitem [{\citenamefont {Cao}\ \emph {et~al.}(2018{\natexlab{a}})\citenamefont
  {Cao}, \citenamefont {Fatemi}, \citenamefont {Fang}, \citenamefont
  {Watanabe}, \citenamefont {Taniguchi}, \citenamefont {Kaxiras},\ and\
  \citenamefont {Jarillo-Herrero}}]{Cao2018a}%
  \BibitemOpen
  \bibfield  {author} {\bibinfo {author} {\bibfnamefont {Y.}~\bibnamefont
  {Cao}}, \bibinfo {author} {\bibfnamefont {V.}~\bibnamefont {Fatemi}},
  \bibinfo {author} {\bibfnamefont {S.}~\bibnamefont {Fang}}, \bibinfo {author}
  {\bibfnamefont {K.}~\bibnamefont {Watanabe}}, \bibinfo {author}
  {\bibfnamefont {T.}~\bibnamefont {Taniguchi}}, \bibinfo {author}
  {\bibfnamefont {E.}~\bibnamefont {Kaxiras}}, \ and\ \bibinfo {author}
  {\bibfnamefont {P.}~\bibnamefont {Jarillo-Herrero}},\ }\href {\doibase
  10.1038/nature26160} {\bibfield  {journal} {\bibinfo  {journal} {Nature}\
  }\textbf {\bibinfo {volume} {556}},\ \bibinfo {pages} {43} (\bibinfo {year}
  {2018}{\natexlab{a}})}\BibitemShut {NoStop}%
\bibitem [{\citenamefont {Cao}\ \emph {et~al.}(2018{\natexlab{b}})\citenamefont
  {Cao}, \citenamefont {Fatemi}, \citenamefont {Demir}, \citenamefont {Fang},
  \citenamefont {Tomarken}, \citenamefont {Luo}, \citenamefont
  {Sanchez-Yamagishi}, \citenamefont {Watanabe}, \citenamefont {Taniguchi},
  \citenamefont {Kaxiras}, \citenamefont {Ashoori},\ and\ \citenamefont
  {Jarillo-Herrero}}]{Cao2018}%
  \BibitemOpen
  \bibfield  {author} {\bibinfo {author} {\bibfnamefont {Y.}~\bibnamefont
  {Cao}}, \bibinfo {author} {\bibfnamefont {V.}~\bibnamefont {Fatemi}},
  \bibinfo {author} {\bibfnamefont {A.}~\bibnamefont {Demir}}, \bibinfo
  {author} {\bibfnamefont {S.}~\bibnamefont {Fang}}, \bibinfo {author}
  {\bibfnamefont {S.~L.}\ \bibnamefont {Tomarken}}, \bibinfo {author}
  {\bibfnamefont {J.~Y.}\ \bibnamefont {Luo}}, \bibinfo {author} {\bibfnamefont
  {J.~D.}\ \bibnamefont {Sanchez-Yamagishi}}, \bibinfo {author} {\bibfnamefont
  {K.}~\bibnamefont {Watanabe}}, \bibinfo {author} {\bibfnamefont
  {T.}~\bibnamefont {Taniguchi}}, \bibinfo {author} {\bibfnamefont
  {E.}~\bibnamefont {Kaxiras}}, \bibinfo {author} {\bibfnamefont {R.~C.}\
  \bibnamefont {Ashoori}}, \ and\ \bibinfo {author} {\bibfnamefont
  {P.}~\bibnamefont {Jarillo-Herrero}},\ }\href {\doibase 10.1038/nature26154}
  {\bibfield  {journal} {\bibinfo  {journal} {Nature}\ }\textbf {\bibinfo
  {volume} {556}},\ \bibinfo {pages} {80} (\bibinfo {year}
  {2018}{\natexlab{b}})}\BibitemShut {NoStop}%
\bibitem [{\citenamefont {Yankowitz}\ \emph {et~al.}(2019)\citenamefont
  {Yankowitz}, \citenamefont {Chen}, \citenamefont {Polshyn}, \citenamefont
  {Zhang}, \citenamefont {Watanabe}, \citenamefont {Taniguchi}, \citenamefont
  {Graf}, \citenamefont {Young},\ and\ \citenamefont {Dean}}]{Yankowitz2019}%
  \BibitemOpen
  \bibfield  {author} {\bibinfo {author} {\bibfnamefont {M.}~\bibnamefont
  {Yankowitz}}, \bibinfo {author} {\bibfnamefont {S.}~\bibnamefont {Chen}},
  \bibinfo {author} {\bibfnamefont {H.}~\bibnamefont {Polshyn}}, \bibinfo
  {author} {\bibfnamefont {Y.}~\bibnamefont {Zhang}}, \bibinfo {author}
  {\bibfnamefont {K.}~\bibnamefont {Watanabe}}, \bibinfo {author}
  {\bibfnamefont {T.}~\bibnamefont {Taniguchi}}, \bibinfo {author}
  {\bibfnamefont {D.}~\bibnamefont {Graf}}, \bibinfo {author} {\bibfnamefont
  {A.~F.}\ \bibnamefont {Young}}, \ and\ \bibinfo {author} {\bibfnamefont
  {C.~R.}\ \bibnamefont {Dean}},\ }\href {\doibase 10.1126/science.aav1910}
  {\bibfield  {journal} {\bibinfo  {journal} {Science}\ }\textbf {\bibinfo
  {volume} {363}},\ \bibinfo {pages} {1059} (\bibinfo {year}
  {2019})}\BibitemShut {NoStop}%
\bibitem [{\citenamefont {Tang}\ \emph {et~al.}(2020)\citenamefont {Tang},
  \citenamefont {Li}, \citenamefont {Li}, \citenamefont {Xu}, \citenamefont
  {Liu}, \citenamefont {Barmak}, \citenamefont {Watanabe}, \citenamefont
  {Taniguchi}, \citenamefont {MacDonald}, \citenamefont {Shan},\ and\
  \citenamefont {Mak}}]{Tang2020}%
  \BibitemOpen
  \bibfield  {author} {\bibinfo {author} {\bibfnamefont {Y.}~\bibnamefont
  {Tang}}, \bibinfo {author} {\bibfnamefont {L.}~\bibnamefont {Li}}, \bibinfo
  {author} {\bibfnamefont {T.}~\bibnamefont {Li}}, \bibinfo {author}
  {\bibfnamefont {Y.}~\bibnamefont {Xu}}, \bibinfo {author} {\bibfnamefont
  {S.}~\bibnamefont {Liu}}, \bibinfo {author} {\bibfnamefont {K.}~\bibnamefont
  {Barmak}}, \bibinfo {author} {\bibfnamefont {K.}~\bibnamefont {Watanabe}},
  \bibinfo {author} {\bibfnamefont {T.}~\bibnamefont {Taniguchi}}, \bibinfo
  {author} {\bibfnamefont {A.~H.}\ \bibnamefont {MacDonald}}, \bibinfo {author}
  {\bibfnamefont {J.}~\bibnamefont {Shan}}, \ and\ \bibinfo {author}
  {\bibfnamefont {K.~F.}\ \bibnamefont {Mak}},\ }\href {\doibase
  10.1038/s41586-020-2085-3} {\bibfield  {journal} {\bibinfo  {journal}
  {Nature}\ }\textbf {\bibinfo {volume} {579}},\ \bibinfo {pages} {353}
  (\bibinfo {year} {2020})}\BibitemShut {NoStop}%
\bibitem [{\citenamefont {Regan}\ \emph {et~al.}(2020)\citenamefont {Regan},
  \citenamefont {Wang}, \citenamefont {Jin}, \citenamefont {{Bakti Utama}},
  \citenamefont {Gao}, \citenamefont {Wei}, \citenamefont {Zhao}, \citenamefont
  {Zhao}, \citenamefont {Zhang}, \citenamefont {Yumigeta}, \citenamefont
  {Blei}, \citenamefont {Carlstr{\"{o}}m}, \citenamefont {Watanabe},
  \citenamefont {Taniguchi}, \citenamefont {Tongay}, \citenamefont {Crommie},
  \citenamefont {Zettl},\ and\ \citenamefont {Wang}}]{Regan2020}%
  \BibitemOpen
  \bibfield  {author} {\bibinfo {author} {\bibfnamefont {E.~C.}\ \bibnamefont
  {Regan}}, \bibinfo {author} {\bibfnamefont {D.}~\bibnamefont {Wang}},
  \bibinfo {author} {\bibfnamefont {C.}~\bibnamefont {Jin}}, \bibinfo {author}
  {\bibfnamefont {M.~I.}\ \bibnamefont {{Bakti Utama}}}, \bibinfo {author}
  {\bibfnamefont {B.}~\bibnamefont {Gao}}, \bibinfo {author} {\bibfnamefont
  {X.}~\bibnamefont {Wei}}, \bibinfo {author} {\bibfnamefont {S.}~\bibnamefont
  {Zhao}}, \bibinfo {author} {\bibfnamefont {W.}~\bibnamefont {Zhao}}, \bibinfo
  {author} {\bibfnamefont {Z.}~\bibnamefont {Zhang}}, \bibinfo {author}
  {\bibfnamefont {K.}~\bibnamefont {Yumigeta}}, \bibinfo {author}
  {\bibfnamefont {M.}~\bibnamefont {Blei}}, \bibinfo {author} {\bibfnamefont
  {J.~D.}\ \bibnamefont {Carlstr{\"{o}}m}}, \bibinfo {author} {\bibfnamefont
  {K.}~\bibnamefont {Watanabe}}, \bibinfo {author} {\bibfnamefont
  {T.}~\bibnamefont {Taniguchi}}, \bibinfo {author} {\bibfnamefont
  {S.}~\bibnamefont {Tongay}}, \bibinfo {author} {\bibfnamefont
  {M.}~\bibnamefont {Crommie}}, \bibinfo {author} {\bibfnamefont
  {A.}~\bibnamefont {Zettl}}, \ and\ \bibinfo {author} {\bibfnamefont
  {F.}~\bibnamefont {Wang}},\ }\href {\doibase 10.1038/s41586-020-2092-4}
  {\bibfield  {journal} {\bibinfo  {journal} {Nature}\ }\textbf {\bibinfo
  {volume} {579}},\ \bibinfo {pages} {359} (\bibinfo {year}
  {2020})}\BibitemShut {NoStop}%
\bibitem [{\citenamefont {Shimazaki}\ \emph {et~al.}(2020)\citenamefont
  {Shimazaki}, \citenamefont {Schwartz}, \citenamefont {Watanabe},
  \citenamefont {Taniguchi}, \citenamefont {Kroner},\ and\ \citenamefont
  {Imamoğlu}}]{Shimazaki2020}%
  \BibitemOpen
  \bibfield  {author} {\bibinfo {author} {\bibfnamefont {Y.}~\bibnamefont
  {Shimazaki}}, \bibinfo {author} {\bibfnamefont {I.}~\bibnamefont {Schwartz}},
  \bibinfo {author} {\bibfnamefont {K.}~\bibnamefont {Watanabe}}, \bibinfo
  {author} {\bibfnamefont {T.}~\bibnamefont {Taniguchi}}, \bibinfo {author}
  {\bibfnamefont {M.}~\bibnamefont {Kroner}}, \ and\ \bibinfo {author}
  {\bibfnamefont {A.}~\bibnamefont {Imamoğlu}},\ }\href {\doibase
  10.1038/s41586-020-2191-2} {\bibfield  {journal} {\bibinfo  {journal}
  {Nature}\ }\textbf {\bibinfo {volume} {580}},\ \bibinfo {pages} {472}
  (\bibinfo {year} {2020})}\BibitemShut {NoStop}%
\bibitem [{\citenamefont {Wang}\ \emph {et~al.}(2020)\citenamefont {Wang},
  \citenamefont {Shih}, \citenamefont {Ghiotto}, \citenamefont {Xian},
  \citenamefont {Rhodes}, \citenamefont {Tan}, \citenamefont {Claassen},
  \citenamefont {Kennes}, \citenamefont {Bai}, \citenamefont {Kim},
  \citenamefont {Watanabe}, \citenamefont {Taniguchi}, \citenamefont {Zhu},
  \citenamefont {Hone}, \citenamefont {Rubio}, \citenamefont {Pasupathy},\ and\
  \citenamefont {Dean}}]{Wang2020}%
  \BibitemOpen
  \bibfield  {author} {\bibinfo {author} {\bibfnamefont {L.}~\bibnamefont
  {Wang}}, \bibinfo {author} {\bibfnamefont {E.-M.}\ \bibnamefont {Shih}},
  \bibinfo {author} {\bibfnamefont {A.}~\bibnamefont {Ghiotto}}, \bibinfo
  {author} {\bibfnamefont {L.}~\bibnamefont {Xian}}, \bibinfo {author}
  {\bibfnamefont {D.~A.}\ \bibnamefont {Rhodes}}, \bibinfo {author}
  {\bibfnamefont {C.}~\bibnamefont {Tan}}, \bibinfo {author} {\bibfnamefont
  {M.}~\bibnamefont {Claassen}}, \bibinfo {author} {\bibfnamefont {D.~M.}\
  \bibnamefont {Kennes}}, \bibinfo {author} {\bibfnamefont {Y.}~\bibnamefont
  {Bai}}, \bibinfo {author} {\bibfnamefont {B.}~\bibnamefont {Kim}}, \bibinfo
  {author} {\bibfnamefont {K.}~\bibnamefont {Watanabe}}, \bibinfo {author}
  {\bibfnamefont {T.}~\bibnamefont {Taniguchi}}, \bibinfo {author}
  {\bibfnamefont {X.}~\bibnamefont {Zhu}}, \bibinfo {author} {\bibfnamefont
  {J.}~\bibnamefont {Hone}}, \bibinfo {author} {\bibfnamefont {A.}~\bibnamefont
  {Rubio}}, \bibinfo {author} {\bibfnamefont {A.~N.}\ \bibnamefont
  {Pasupathy}}, \ and\ \bibinfo {author} {\bibfnamefont {C.~R.}\ \bibnamefont
  {Dean}},\ }\href {\doibase 10.1038/s41563-020-0708-6} {\bibfield  {journal}
  {\bibinfo  {journal} {Nature Materials}\ }\textbf {\bibinfo {volume} {19}},\
  \bibinfo {pages} {861} (\bibinfo {year} {2020})}\BibitemShut {NoStop}%
\bibitem [{\citenamefont {Xu}\ \emph {et~al.}(2020)\citenamefont {Xu},
  \citenamefont {Liu}, \citenamefont {Rhodes}, \citenamefont {Watanabe},
  \citenamefont {Taniguchi}, \citenamefont {Hone}, \citenamefont {Elser},
  \citenamefont {Mak},\ and\ \citenamefont {Shan}}]{Xu2020}%
  \BibitemOpen
  \bibfield  {author} {\bibinfo {author} {\bibfnamefont {Y.}~\bibnamefont
  {Xu}}, \bibinfo {author} {\bibfnamefont {S.}~\bibnamefont {Liu}}, \bibinfo
  {author} {\bibfnamefont {D.~A.}\ \bibnamefont {Rhodes}}, \bibinfo {author}
  {\bibfnamefont {K.}~\bibnamefont {Watanabe}}, \bibinfo {author}
  {\bibfnamefont {T.}~\bibnamefont {Taniguchi}}, \bibinfo {author}
  {\bibfnamefont {J.}~\bibnamefont {Hone}}, \bibinfo {author} {\bibfnamefont
  {V.}~\bibnamefont {Elser}}, \bibinfo {author} {\bibfnamefont {K.~F.}\
  \bibnamefont {Mak}}, \ and\ \bibinfo {author} {\bibfnamefont
  {J.}~\bibnamefont {Shan}},\ }\href {http://arxiv.org/abs/2007.11128}\ \Eprint {http://arxiv.org/abs/2007.11128}
  {arXiv:2007.11128} \BibitemShut {NoStop}%
\bibitem [{\citenamefont {Xu}\ \emph {et~al.}(2014)\citenamefont {Xu},
  \citenamefont {Yao}, \citenamefont {Xiao},\ and\ \citenamefont
  {Heinz}}]{Xu2014}%
  \BibitemOpen
  \bibfield  {author} {\bibinfo {author} {\bibfnamefont {X.}~\bibnamefont
  {Xu}}, \bibinfo {author} {\bibfnamefont {W.}~\bibnamefont {Yao}}, \bibinfo
  {author} {\bibfnamefont {D.}~\bibnamefont {Xiao}}, \ and\ \bibinfo {author}
  {\bibfnamefont {T.~F.}\ \bibnamefont {Heinz}},\ }\href {\doibase
  10.1038/nphys2942} {\bibfield  {journal} {\bibinfo  {journal} {Nature
  Physics}\ }\textbf {\bibinfo {volume} {10}},\ \bibinfo {pages} {343}
  (\bibinfo {year} {2014})}\BibitemShut {NoStop}%
\bibitem [{\citenamefont {Wang}\ \emph {et~al.}(2018)\citenamefont {Wang},
  \citenamefont {Chernikov}, \citenamefont {Glazov}, \citenamefont {Heinz},
  \citenamefont {Marie}, \citenamefont {Amand},\ and\ \citenamefont
  {Urbaszek}}]{Wang2018}%
  \BibitemOpen
  \bibfield  {author} {\bibinfo {author} {\bibfnamefont {G.}~\bibnamefont
  {Wang}}, \bibinfo {author} {\bibfnamefont {A.}~\bibnamefont {Chernikov}},
  \bibinfo {author} {\bibfnamefont {M.~M.}\ \bibnamefont {Glazov}}, \bibinfo
  {author} {\bibfnamefont {T.~F.}\ \bibnamefont {Heinz}}, \bibinfo {author}
  {\bibfnamefont {X.}~\bibnamefont {Marie}}, \bibinfo {author} {\bibfnamefont
  {T.}~\bibnamefont {Amand}}, \ and\ \bibinfo {author} {\bibfnamefont
  {B.}~\bibnamefont {Urbaszek}},\ }\href {\doibase
  10.1103/RevModPhys.90.021001} {\bibfield  {journal} {\bibinfo  {journal}
  {Reviews of Modern Physics}\ }\textbf {\bibinfo {volume} {90}},\ \bibinfo
  {pages} {021001} (\bibinfo {year} {2018})}\BibitemShut {NoStop}%
\bibitem [{\citenamefont {Fey}\ \emph {et~al.}(2020)\citenamefont {Fey},
  \citenamefont {Schmelcher}, \citenamefont {Imamoglu},\ and\ \citenamefont
  {Schmidt}}]{Fey2020}%
  \BibitemOpen
  \bibfield  {author} {\bibinfo {author} {\bibfnamefont {C.}~\bibnamefont
  {Fey}}, \bibinfo {author} {\bibfnamefont {P.}~\bibnamefont {Schmelcher}},
  \bibinfo {author} {\bibfnamefont {A.}~\bibnamefont {Imamoglu}}, \ and\
  \bibinfo {author} {\bibfnamefont {R.}~\bibnamefont {Schmidt}},\ }\href
  {\doibase 10.1103/PhysRevB.101.195417} {\bibfield  {journal} {\bibinfo
  {journal} {Physical Review B}\ }\textbf {\bibinfo {volume} {101}},\ \bibinfo
  {pages} {195417} (\bibinfo {year} {2020})}\BibitemShut {NoStop}%
\bibitem [{\citenamefont {Camjayi}\ \emph {et~al.}(2008)\citenamefont
  {Camjayi}, \citenamefont {Haule}, \citenamefont {Dobrosavljevi{\'{c}}},\ and\
  \citenamefont {Kotliar}}]{Camjayi2008}%
  \BibitemOpen
  \bibfield  {author} {\bibinfo {author} {\bibfnamefont {A.}~\bibnamefont
  {Camjayi}}, \bibinfo {author} {\bibfnamefont {K.}~\bibnamefont {Haule}},
  \bibinfo {author} {\bibfnamefont {V.}~\bibnamefont {Dobrosavljevi{\'{c}}}}, \
  and\ \bibinfo {author} {\bibfnamefont {G.}~\bibnamefont {Kotliar}},\ }\href
  {\doibase 10.1038/nphys1106} {\bibfield  {journal} {\bibinfo  {journal}
  {Nature Physics}\ }\textbf {\bibinfo {volume} {4}},\ \bibinfo {pages} {932}
  (\bibinfo {year} {2008})}\BibitemShut {NoStop}%
\bibitem [{\citenamefont {Sidler}\ \emph {et~al.}(2017)\citenamefont {Sidler},
  \citenamefont {Back}, \citenamefont {Cotlet}, \citenamefont {Srivastava},
  \citenamefont {Fink}, \citenamefont {Kroner}, \citenamefont {Demler},\ and\
  \citenamefont {Imamoglu}}]{Sidler2017}%
  \BibitemOpen
  \bibfield  {author} {\bibinfo {author} {\bibfnamefont {M.}~\bibnamefont
  {Sidler}}, \bibinfo {author} {\bibfnamefont {P.}~\bibnamefont {Back}},
  \bibinfo {author} {\bibfnamefont {O.}~\bibnamefont {Cotlet}}, \bibinfo
  {author} {\bibfnamefont {A.}~\bibnamefont {Srivastava}}, \bibinfo {author}
  {\bibfnamefont {T.}~\bibnamefont {Fink}}, \bibinfo {author} {\bibfnamefont
  {M.}~\bibnamefont {Kroner}}, \bibinfo {author} {\bibfnamefont
  {E.}~\bibnamefont {Demler}}, \ and\ \bibinfo {author} {\bibfnamefont
  {A.}~\bibnamefont {Imamoglu}},\ }\href {\doibase 10.1038/nphys3949}
  {\bibfield  {journal} {\bibinfo  {journal} {Nature Physics}\ }\textbf
  {\bibinfo {volume} {13}},\ \bibinfo {pages} {255} (\bibinfo {year}
  {2017})}\BibitemShut {NoStop}%
\bibitem [{\citenamefont {Efimkin}\ and\ \citenamefont
  {MacDonald}(2017)}]{Efimkin2017}%
  \BibitemOpen
  \bibfield  {author} {\bibinfo {author} {\bibfnamefont {D.~K.}\ \bibnamefont
  {Efimkin}}\ and\ \bibinfo {author} {\bibfnamefont {A.~H.}\ \bibnamefont
  {MacDonald}},\ }\href {\doibase 10.1103/PhysRevB.95.035417} {\bibfield
  {journal} {\bibinfo  {journal} {Physical Review B}\ }\textbf {\bibinfo
  {volume} {95}},\ \bibinfo {pages} {035417} (\bibinfo {year}
  {2017})}\BibitemShut {NoStop}%
\bibitem [{\citenamefont {Seyler}\ \emph {et~al.}(2019)\citenamefont {Seyler},
  \citenamefont {Rivera}, \citenamefont {Yu}, \citenamefont {Wilson},
  \citenamefont {Ray}, \citenamefont {Mandrus}, \citenamefont {Yan},
  \citenamefont {Yao},\ and\ \citenamefont {Xu}}]{Seyler2019}%
  \BibitemOpen
  \bibfield  {author} {\bibinfo {author} {\bibfnamefont {K.~L.}\ \bibnamefont
  {Seyler}}, \bibinfo {author} {\bibfnamefont {P.}~\bibnamefont {Rivera}},
  \bibinfo {author} {\bibfnamefont {H.}~\bibnamefont {Yu}}, \bibinfo {author}
  {\bibfnamefont {N.~P.}\ \bibnamefont {Wilson}}, \bibinfo {author}
  {\bibfnamefont {E.~L.}\ \bibnamefont {Ray}}, \bibinfo {author} {\bibfnamefont
  {D.~G.}\ \bibnamefont {Mandrus}}, \bibinfo {author} {\bibfnamefont
  {J.}~\bibnamefont {Yan}}, \bibinfo {author} {\bibfnamefont {W.}~\bibnamefont
  {Yao}}, \ and\ \bibinfo {author} {\bibfnamefont {X.}~\bibnamefont {Xu}},\
  }\href {\doibase 10.1038/s41586-019-0957-1} {\bibfield  {journal} {\bibinfo
  {journal} {Nature}\ }\textbf {\bibinfo {volume} {567}},\ \bibinfo {pages}
  {66} (\bibinfo {year} {2019})}\BibitemShut {NoStop}%
\bibitem [{\citenamefont {Tran}\ \emph {et~al.}(2019)\citenamefont {Tran},
  \citenamefont {Moody}, \citenamefont {Wu}, \citenamefont {Lu}, \citenamefont
  {Choi}, \citenamefont {Kim}, \citenamefont {Rai}, \citenamefont {Sanchez},
  \citenamefont {Quan}, \citenamefont {Singh}, \citenamefont {Embley},
  \citenamefont {Zepeda}, \citenamefont {Campbell}, \citenamefont {Autry},
  \citenamefont {Taniguchi}, \citenamefont {Watanabe}, \citenamefont {Lu},
  \citenamefont {Banerjee}, \citenamefont {Silverman}, \citenamefont {Kim},
  \citenamefont {Tutuc}, \citenamefont {Yang}, \citenamefont {MacDonald},\ and\
  \citenamefont {Li}}]{Tran2019}%
  \BibitemOpen
  \bibfield  {author} {\bibinfo {author} {\bibfnamefont {K.}~\bibnamefont
  {Tran}}, \bibinfo {author} {\bibfnamefont {G.}~\bibnamefont {Moody}},
  \bibinfo {author} {\bibfnamefont {F.}~\bibnamefont {Wu}}, \bibinfo {author}
  {\bibfnamefont {X.}~\bibnamefont {Lu}}, \bibinfo {author} {\bibfnamefont
  {J.}~\bibnamefont {Choi}}, \bibinfo {author} {\bibfnamefont {K.}~\bibnamefont
  {Kim}}, \bibinfo {author} {\bibfnamefont {A.}~\bibnamefont {Rai}}, \bibinfo
  {author} {\bibfnamefont {D.~A.}\ \bibnamefont {Sanchez}}, \bibinfo {author}
  {\bibfnamefont {J.}~\bibnamefont {Quan}}, \bibinfo {author} {\bibfnamefont
  {A.}~\bibnamefont {Singh}}, \bibinfo {author} {\bibfnamefont
  {J.}~\bibnamefont {Embley}}, \bibinfo {author} {\bibfnamefont
  {A.}~\bibnamefont {Zepeda}}, \bibinfo {author} {\bibfnamefont
  {M.}~\bibnamefont {Campbell}}, \bibinfo {author} {\bibfnamefont
  {T.}~\bibnamefont {Autry}}, \bibinfo {author} {\bibfnamefont
  {T.}~\bibnamefont {Taniguchi}}, \bibinfo {author} {\bibfnamefont
  {K.}~\bibnamefont {Watanabe}}, \bibinfo {author} {\bibfnamefont
  {N.}~\bibnamefont {Lu}}, \bibinfo {author} {\bibfnamefont {S.~K.}\
  \bibnamefont {Banerjee}}, \bibinfo {author} {\bibfnamefont {K.~L.}\
  \bibnamefont {Silverman}}, \bibinfo {author} {\bibfnamefont {S.}~\bibnamefont
  {Kim}}, \bibinfo {author} {\bibfnamefont {E.}~\bibnamefont {Tutuc}}, \bibinfo
  {author} {\bibfnamefont {L.}~\bibnamefont {Yang}}, \bibinfo {author}
  {\bibfnamefont {A.~H.}\ \bibnamefont {MacDonald}}, \ and\ \bibinfo {author}
  {\bibfnamefont {X.}~\bibnamefont {Li}},\ }\href {\doibase
  10.1038/s41586-019-0975-z} {\bibfield  {journal} {\bibinfo  {journal}
  {Nature}\ }\textbf {\bibinfo {volume} {567}},\ \bibinfo {pages} {71}
  (\bibinfo {year} {2019})}\BibitemShut {NoStop}%
\bibitem [{\citenamefont {Jin}\ \emph {et~al.}(2019)\citenamefont {Jin},
  \citenamefont {Regan}, \citenamefont {Yan}, \citenamefont {{Iqbal Bakti
  Utama}}, \citenamefont {Wang}, \citenamefont {Zhao}, \citenamefont {Qin},
  \citenamefont {Yang}, \citenamefont {Zheng}, \citenamefont {Shi},
  \citenamefont {Watanabe}, \citenamefont {Taniguchi}, \citenamefont {Tongay},
  \citenamefont {Zettl},\ and\ \citenamefont {Wang}}]{Jin2019}%
  \BibitemOpen
  \bibfield  {author} {\bibinfo {author} {\bibfnamefont {C.}~\bibnamefont
  {Jin}}, \bibinfo {author} {\bibfnamefont {E.~C.}\ \bibnamefont {Regan}},
  \bibinfo {author} {\bibfnamefont {A.}~\bibnamefont {Yan}}, \bibinfo {author}
  {\bibfnamefont {M.}~\bibnamefont {{Iqbal Bakti Utama}}}, \bibinfo {author}
  {\bibfnamefont {D.}~\bibnamefont {Wang}}, \bibinfo {author} {\bibfnamefont
  {S.}~\bibnamefont {Zhao}}, \bibinfo {author} {\bibfnamefont {Y.}~\bibnamefont
  {Qin}}, \bibinfo {author} {\bibfnamefont {S.}~\bibnamefont {Yang}}, \bibinfo
  {author} {\bibfnamefont {Z.}~\bibnamefont {Zheng}}, \bibinfo {author}
  {\bibfnamefont {S.}~\bibnamefont {Shi}}, \bibinfo {author} {\bibfnamefont
  {K.}~\bibnamefont {Watanabe}}, \bibinfo {author} {\bibfnamefont
  {T.}~\bibnamefont {Taniguchi}}, \bibinfo {author} {\bibfnamefont
  {S.}~\bibnamefont {Tongay}}, \bibinfo {author} {\bibfnamefont
  {A.}~\bibnamefont {Zettl}}, \ and\ \bibinfo {author} {\bibfnamefont
  {F.}~\bibnamefont {Wang}},\ }\href {\doibase 10.1038/s41586-019-0976-y}
  {\bibfield  {journal} {\bibinfo  {journal} {Nature}\ }\textbf {\bibinfo
  {volume} {567}},\ \bibinfo {pages} {76} (\bibinfo {year} {2019})}\BibitemShut
  {NoStop}%
\bibitem [{\citenamefont {Alexeev}\ \emph {et~al.}(2019)\citenamefont
  {Alexeev}, \citenamefont {Ruiz-Tijerina}, \citenamefont {Danovich},
  \citenamefont {Hamer}, \citenamefont {Terry}, \citenamefont {Nayak},
  \citenamefont {Ahn}, \citenamefont {Pak}, \citenamefont {Lee}, \citenamefont
  {Sohn}, \citenamefont {Molas}, \citenamefont {Koperski}, \citenamefont
  {Watanabe}, \citenamefont {Taniguchi}, \citenamefont {Novoselov},
  \citenamefont {Gorbachev}, \citenamefont {Shin}, \citenamefont {Fal'ko},\
  and\ \citenamefont {Tartakovskii}}]{Alexeev2019}%
  \BibitemOpen
  \bibfield  {author} {\bibinfo {author} {\bibfnamefont {E.~M.}\ \bibnamefont
  {Alexeev}}, \bibinfo {author} {\bibfnamefont {D.~A.}\ \bibnamefont
  {Ruiz-Tijerina}}, \bibinfo {author} {\bibfnamefont {M.}~\bibnamefont
  {Danovich}}, \bibinfo {author} {\bibfnamefont {M.~J.}\ \bibnamefont {Hamer}},
  \bibinfo {author} {\bibfnamefont {D.~J.}\ \bibnamefont {Terry}}, \bibinfo
  {author} {\bibfnamefont {P.~K.}\ \bibnamefont {Nayak}}, \bibinfo {author}
  {\bibfnamefont {S.}~\bibnamefont {Ahn}}, \bibinfo {author} {\bibfnamefont
  {S.}~\bibnamefont {Pak}}, \bibinfo {author} {\bibfnamefont {J.}~\bibnamefont
  {Lee}}, \bibinfo {author} {\bibfnamefont {J.~I.}\ \bibnamefont {Sohn}},
  \bibinfo {author} {\bibfnamefont {M.~R.}\ \bibnamefont {Molas}}, \bibinfo
  {author} {\bibfnamefont {M.}~\bibnamefont {Koperski}}, \bibinfo {author}
  {\bibfnamefont {K.}~\bibnamefont {Watanabe}}, \bibinfo {author}
  {\bibfnamefont {T.}~\bibnamefont {Taniguchi}}, \bibinfo {author}
  {\bibfnamefont {K.~S.}\ \bibnamefont {Novoselov}}, \bibinfo {author}
  {\bibfnamefont {R.~V.}\ \bibnamefont {Gorbachev}}, \bibinfo {author}
  {\bibfnamefont {H.~S.}\ \bibnamefont {Shin}}, \bibinfo {author}
  {\bibfnamefont {V.~I.}\ \bibnamefont {Fal'ko}}, \ and\ \bibinfo {author}
  {\bibfnamefont {A.~I.}\ \bibnamefont {Tartakovskii}},\ }\href {\doibase
  10.1038/s41586-019-0986-9} {\bibfield  {journal} {\bibinfo  {journal}
  {Nature}\ }\textbf {\bibinfo {volume} {567}},\ \bibinfo {pages} {81}
  (\bibinfo {year} {2019})}\BibitemShut {NoStop}%
\bibitem [{\citenamefont {Andersen}\ \emph {et~al.}(2019)\citenamefont
  {Andersen}, \citenamefont {Scuri}, \citenamefont {Sushko}, \citenamefont {{De
  Greve}}, \citenamefont {Sung}, \citenamefont {Zhou}, \citenamefont {Wild},
  \citenamefont {Gelly}, \citenamefont {Heo}, \citenamefont {Watanabe},
  \citenamefont {Taniguchi}, \citenamefont {Kim}, \citenamefont {Park},\ and\
  \citenamefont {Lukin}}]{Andersen2019}%
  \BibitemOpen
  \bibfield  {author} {\bibinfo {author} {\bibfnamefont {T.~I.}\ \bibnamefont
  {Andersen}}, \bibinfo {author} {\bibfnamefont {G.}~\bibnamefont {Scuri}},
  \bibinfo {author} {\bibfnamefont {A.}~\bibnamefont {Sushko}}, \bibinfo
  {author} {\bibfnamefont {K.}~\bibnamefont {{De Greve}}}, \bibinfo {author}
  {\bibfnamefont {J.}~\bibnamefont {Sung}}, \bibinfo {author} {\bibfnamefont
  {Y.}~\bibnamefont {Zhou}}, \bibinfo {author} {\bibfnamefont {D.~S.}\
  \bibnamefont {Wild}}, \bibinfo {author} {\bibfnamefont {R.~J.}\ \bibnamefont
  {Gelly}}, \bibinfo {author} {\bibfnamefont {H.}~\bibnamefont {Heo}}, \bibinfo
  {author} {\bibfnamefont {K.}~\bibnamefont {Watanabe}}, \bibinfo {author}
  {\bibfnamefont {T.}~\bibnamefont {Taniguchi}}, \bibinfo {author}
  {\bibfnamefont {P.}~\bibnamefont {Kim}}, \bibinfo {author} {\bibfnamefont
  {H.}~\bibnamefont {Park}}, \ and\ \bibinfo {author} {\bibfnamefont {M.~D.}\
  \bibnamefont {Lukin}},\ }\href {https://arxiv.org/abs/1912.06955} \ \Eprint {http://arxiv.org/abs/1912.06955}
  {arXiv:1912.06955} \BibitemShut {NoStop}%
\bibitem [{\citenamefont {Yu}\ \emph {et~al.}(2017)\citenamefont {Yu},
  \citenamefont {Liu}, \citenamefont {Tang}, \citenamefont {Xu},\ and\
  \citenamefont {Yao}}]{Yu2017}%
  \BibitemOpen
  \bibfield  {author} {\bibinfo {author} {\bibfnamefont {H.}~\bibnamefont
  {Yu}}, \bibinfo {author} {\bibfnamefont {G.-B.}\ \bibnamefont {Liu}},
  \bibinfo {author} {\bibfnamefont {J.}~\bibnamefont {Tang}}, \bibinfo {author}
  {\bibfnamefont {X.}~\bibnamefont {Xu}}, \ and\ \bibinfo {author}
  {\bibfnamefont {W.}~\bibnamefont {Yao}},\ }\href {\doibase
  10.1126/sciadv.1701696} {\bibfield  {journal} {\bibinfo  {journal} {Science
  Advances}\ }\textbf {\bibinfo {volume} {3}},\ \bibinfo {pages} {e1701696}
  (\bibinfo {year} {2017})}\BibitemShut {NoStop}%
\bibitem [{\citenamefont {Wu}\ \emph {et~al.}(2018)\citenamefont {Wu},
  \citenamefont {Lovorn},\ and\ \citenamefont {MacDonald}}]{Wu2018a}%
  \BibitemOpen
  \bibfield  {author} {\bibinfo {author} {\bibfnamefont {F.}~\bibnamefont
  {Wu}}, \bibinfo {author} {\bibfnamefont {T.}~\bibnamefont {Lovorn}}, \ and\
  \bibinfo {author} {\bibfnamefont {A.~H.}\ \bibnamefont {MacDonald}},\ }\href
  {\doibase 10.1103/PhysRevB.97.035306} {\bibfield  {journal} {\bibinfo
  {journal} {Physical Review B}\ }\textbf {\bibinfo {volume} {97}},\ \bibinfo
  {pages} {035306} (\bibinfo {year} {2018})}\BibitemShut {NoStop}%
\bibitem [{\citenamefont {Ruiz-Tijerina}\ and\ \citenamefont
  {Fal'ko}(2019)}]{Ruiz-Tijerina2019}%
  \BibitemOpen
  \bibfield  {author} {\bibinfo {author} {\bibfnamefont {D.~A.}\ \bibnamefont
  {Ruiz-Tijerina}}\ and\ \bibinfo {author} {\bibfnamefont {V.~I.}\ \bibnamefont
  {Fal'ko}},\ }\href {\doibase 10.1103/PhysRevB.99.125424} {\bibfield
  {journal} {\bibinfo  {journal} {Physical Review B}\ }\textbf {\bibinfo
  {volume} {99}},\ \bibinfo {pages} {125424} (\bibinfo {year}
  {2019})}\BibitemShut {NoStop}%
\bibitem [{\citenamefont {He}\ \emph {et~al.}(2013)\citenamefont {He},
  \citenamefont {Poole}, \citenamefont {Mak},\ and\ \citenamefont
  {Shan}}]{He2013}%
  \BibitemOpen
  \bibfield  {author} {\bibinfo {author} {\bibfnamefont {K.}~\bibnamefont
  {He}}, \bibinfo {author} {\bibfnamefont {C.}~\bibnamefont {Poole}}, \bibinfo
  {author} {\bibfnamefont {K.~F.}\ \bibnamefont {Mak}}, \ and\ \bibinfo
  {author} {\bibfnamefont {J.}~\bibnamefont {Shan}},\ }\href {\doibase
  10.1021/nl4013166} {\bibfield  {journal} {\bibinfo  {journal} {Nano Letters}\
  }\textbf {\bibinfo {volume} {13}},\ \bibinfo {pages} {2931} (\bibinfo {year}
  {2013})}\BibitemShut {NoStop}%
\bibitem [{\citenamefont {Conley}\ \emph {et~al.}(2013)\citenamefont {Conley},
  \citenamefont {Wang}, \citenamefont {Ziegler}, \citenamefont {Haglund},
  \citenamefont {Pantelides},\ and\ \citenamefont {Bolotin}}]{Conley2013}%
  \BibitemOpen
  \bibfield  {author} {\bibinfo {author} {\bibfnamefont {H.~J.}\ \bibnamefont
  {Conley}}, \bibinfo {author} {\bibfnamefont {B.}~\bibnamefont {Wang}},
  \bibinfo {author} {\bibfnamefont {J.~I.}\ \bibnamefont {Ziegler}}, \bibinfo
  {author} {\bibfnamefont {R.~F.}\ \bibnamefont {Haglund}}, \bibinfo {author}
  {\bibfnamefont {S.~T.}\ \bibnamefont {Pantelides}}, \ and\ \bibinfo {author}
  {\bibfnamefont {K.~I.}\ \bibnamefont {Bolotin}},\ }\href {\doibase
  10.1021/nl4014748} {\bibfield  {journal} {\bibinfo  {journal} {Nano Letters}\
  }\textbf {\bibinfo {volume} {13}},\ \bibinfo {pages} {3626} (\bibinfo {year}
  {2013})}\BibitemShut {NoStop}%
\bibitem [{\citenamefont {Zhu}\ \emph {et~al.}(2013)\citenamefont {Zhu},
  \citenamefont {Wang}, \citenamefont {Liu}, \citenamefont {Marie},
  \citenamefont {Qiao}, \citenamefont {Zhang}, \citenamefont {Wu},
  \citenamefont {Fan}, \citenamefont {Tan}, \citenamefont {Amand},\ and\
  \citenamefont {Urbaszek}}]{Zhu2013}%
  \BibitemOpen
  \bibfield  {author} {\bibinfo {author} {\bibfnamefont {C.~R.}\ \bibnamefont
  {Zhu}}, \bibinfo {author} {\bibfnamefont {G.}~\bibnamefont {Wang}}, \bibinfo
  {author} {\bibfnamefont {B.~L.}\ \bibnamefont {Liu}}, \bibinfo {author}
  {\bibfnamefont {X.}~\bibnamefont {Marie}}, \bibinfo {author} {\bibfnamefont
  {X.~F.}\ \bibnamefont {Qiao}}, \bibinfo {author} {\bibfnamefont
  {X.}~\bibnamefont {Zhang}}, \bibinfo {author} {\bibfnamefont {X.~X.}\
  \bibnamefont {Wu}}, \bibinfo {author} {\bibfnamefont {H.}~\bibnamefont
  {Fan}}, \bibinfo {author} {\bibfnamefont {P.~H.}\ \bibnamefont {Tan}},
  \bibinfo {author} {\bibfnamefont {T.}~\bibnamefont {Amand}}, \ and\ \bibinfo
  {author} {\bibfnamefont {B.}~\bibnamefont {Urbaszek}},\ }\href {\doibase
  10.1103/PhysRevB.88.121301} {\bibfield  {journal} {\bibinfo  {journal}
  {Physical Review B}\ }\textbf {\bibinfo {volume} {88}},\ \bibinfo {pages}
  {121301} (\bibinfo {year} {2013})}\BibitemShut {NoStop}%
\bibitem [{\citenamefont {Yu}\ \emph {et~al.}(2014)\citenamefont {Yu},
  \citenamefont {Liu}, \citenamefont {Gong}, \citenamefont {Xu},\ and\
  \citenamefont {Yao}}]{Yu2014}%
  \BibitemOpen
  \bibfield  {author} {\bibinfo {author} {\bibfnamefont {H.}~\bibnamefont
  {Yu}}, \bibinfo {author} {\bibfnamefont {G.-B.}\ \bibnamefont {Liu}},
  \bibinfo {author} {\bibfnamefont {P.}~\bibnamefont {Gong}}, \bibinfo {author}
  {\bibfnamefont {X.}~\bibnamefont {Xu}}, \ and\ \bibinfo {author}
  {\bibfnamefont {W.}~\bibnamefont {Yao}},\ }\href {\doibase
  10.1038/ncomms4876} {\bibfield  {journal} {\bibinfo  {journal} {Nature
  Communications}\ }\textbf {\bibinfo {volume} {5}},\ \bibinfo {pages} {3876}
  (\bibinfo {year} {2014})}\BibitemShut {NoStop}%
\bibitem [{\citenamefont {Glazov}\ \emph {et~al.}(2014)\citenamefont {Glazov},
  \citenamefont {Amand}, \citenamefont {Marie}, \citenamefont {Lagarde},
  \citenamefont {Bouet},\ and\ \citenamefont {Urbaszek}}]{Glazov2014}%
  \BibitemOpen
  \bibfield  {author} {\bibinfo {author} {\bibfnamefont {M.~M.}\ \bibnamefont
  {Glazov}}, \bibinfo {author} {\bibfnamefont {T.}~\bibnamefont {Amand}},
  \bibinfo {author} {\bibfnamefont {X.}~\bibnamefont {Marie}}, \bibinfo
  {author} {\bibfnamefont {D.}~\bibnamefont {Lagarde}}, \bibinfo {author}
  {\bibfnamefont {L.}~\bibnamefont {Bouet}}, \ and\ \bibinfo {author}
  {\bibfnamefont {B.}~\bibnamefont {Urbaszek}},\ }\href {\doibase
  10.1103/PhysRevB.89.201302} {\bibfield  {journal} {\bibinfo  {journal}
  {Physical Review B}\ }\textbf {\bibinfo {volume} {89}},\ \bibinfo {pages}
  {201302} (\bibinfo {year} {2014})}\BibitemShut {NoStop}%
\bibitem [{\citenamefont {Qiu}\ \emph {et~al.}(2015)\citenamefont {Qiu},
  \citenamefont {Cao},\ and\ \citenamefont {Louie}}]{Qiu2015}%
  \BibitemOpen
  \bibfield  {author} {\bibinfo {author} {\bibfnamefont {D.~Y.}\ \bibnamefont
  {Qiu}}, \bibinfo {author} {\bibfnamefont {T.}~\bibnamefont {Cao}}, \ and\
  \bibinfo {author} {\bibfnamefont {S.~G.}\ \bibnamefont {Louie}},\ }\href
  {\doibase 10.1103/PhysRevLett.115.176801} {\bibfield  {journal} {\bibinfo
  {journal} {Physical Review Letters}\ }\textbf {\bibinfo {volume} {115}},\
  \bibinfo {pages} {176801} (\bibinfo {year} {2015})}\BibitemShut {NoStop}%
\bibitem [{\citenamefont {Larentis}\ \emph {et~al.}(2018)\citenamefont
  {Larentis}, \citenamefont {Movva}, \citenamefont {Fallahazad}, \citenamefont
  {Kim}, \citenamefont {Behroozi}, \citenamefont {Taniguchi}, \citenamefont
  {Watanabe}, \citenamefont {Banerjee},\ and\ \citenamefont
  {Tutuc}}]{Larentis2018}%
  \BibitemOpen
  \bibfield  {author} {\bibinfo {author} {\bibfnamefont {S.}~\bibnamefont
  {Larentis}}, \bibinfo {author} {\bibfnamefont {H.~C.~P.}\ \bibnamefont
  {Movva}}, \bibinfo {author} {\bibfnamefont {B.}~\bibnamefont {Fallahazad}},
  \bibinfo {author} {\bibfnamefont {K.}~\bibnamefont {Kim}}, \bibinfo {author}
  {\bibfnamefont {A.}~\bibnamefont {Behroozi}}, \bibinfo {author}
  {\bibfnamefont {T.}~\bibnamefont {Taniguchi}}, \bibinfo {author}
  {\bibfnamefont {K.}~\bibnamefont {Watanabe}}, \bibinfo {author}
  {\bibfnamefont {S.~K.}\ \bibnamefont {Banerjee}}, \ and\ \bibinfo {author}
  {\bibfnamefont {E.}~\bibnamefont {Tutuc}},\ }\href {\doibase
  10.1103/PhysRevB.97.201407} {\bibfield  {journal} {\bibinfo  {journal}
  {Physical Review B}\ }\textbf {\bibinfo {volume} {97}},\ \bibinfo {pages}
  {201407} (\bibinfo {year} {2018})}\BibitemShut {NoStop}%
\bibitem [{\citenamefont {Zhang}\ \emph {et~al.}(2014)\citenamefont {Zhang},
  \citenamefont {Chang}, \citenamefont {Zhou}, \citenamefont {Cui},
  \citenamefont {Yan}, \citenamefont {Liu}, \citenamefont {Schmitt},
  \citenamefont {Lee}, \citenamefont {Moore}, \citenamefont {Chen},
  \citenamefont {Lin}, \citenamefont {Jeng}, \citenamefont {Mo}, \citenamefont
  {Hussain}, \citenamefont {Bansil},\ and\ \citenamefont {Shen}}]{Zhang2014}%
  \BibitemOpen
  \bibfield  {author} {\bibinfo {author} {\bibfnamefont {Y.}~\bibnamefont
  {Zhang}}, \bibinfo {author} {\bibfnamefont {T.-R.}\ \bibnamefont {Chang}},
  \bibinfo {author} {\bibfnamefont {B.}~\bibnamefont {Zhou}}, \bibinfo {author}
  {\bibfnamefont {Y.-t.}\ \bibnamefont {Cui}}, \bibinfo {author} {\bibfnamefont
  {H.}~\bibnamefont {Yan}}, \bibinfo {author} {\bibfnamefont {Z.}~\bibnamefont
  {Liu}}, \bibinfo {author} {\bibfnamefont {F.}~\bibnamefont {Schmitt}},
  \bibinfo {author} {\bibfnamefont {J.}~\bibnamefont {Lee}}, \bibinfo {author}
  {\bibfnamefont {R.}~\bibnamefont {Moore}}, \bibinfo {author} {\bibfnamefont
  {Y.}~\bibnamefont {Chen}}, \bibinfo {author} {\bibfnamefont {H.}~\bibnamefont
  {Lin}}, \bibinfo {author} {\bibfnamefont {H.-T.}\ \bibnamefont {Jeng}},
  \bibinfo {author} {\bibfnamefont {S.-K.}\ \bibnamefont {Mo}}, \bibinfo
  {author} {\bibfnamefont {Z.}~\bibnamefont {Hussain}}, \bibinfo {author}
  {\bibfnamefont {A.}~\bibnamefont {Bansil}}, \ and\ \bibinfo {author}
  {\bibfnamefont {Z.-X.}\ \bibnamefont {Shen}},\ }\href {\doibase
  10.1038/nnano.2013.277} {\bibfield  {journal} {\bibinfo  {journal} {Nature
  Nanotechnology}\ }\textbf {\bibinfo {volume} {9}},\ \bibinfo {pages} {111}
  (\bibinfo {year} {2014})}\BibitemShut {NoStop}%
\bibitem [{\citenamefont {Yoshioka}\ and\ \citenamefont
  {Fukuyama}(1979)}]{Yoshioka1979}%
  \BibitemOpen
  \bibfield  {author} {\bibinfo {author} {\bibfnamefont {D.}~\bibnamefont
  {Yoshioka}}\ and\ \bibinfo {author} {\bibfnamefont {H.}~\bibnamefont
  {Fukuyama}},\ }\href {\doibase 10.1143/JPSJ.47.394} {\bibfield  {journal}
  {\bibinfo  {journal} {Journal of the Physical Society of Japan}\ }\textbf
  {\bibinfo {volume} {47}},\ \bibinfo {pages} {394} (\bibinfo {year}
  {1979})}\BibitemShut {NoStop}%
\bibitem [{\citenamefont {Li}\ \emph {et~al.}(2014)\citenamefont {Li},
  \citenamefont {Ludwig}, \citenamefont {Low}, \citenamefont {Chernikov},
  \citenamefont {Cui}, \citenamefont {Arefe}, \citenamefont {Kim},
  \citenamefont {van~der Zande}, \citenamefont {Rigosi}, \citenamefont {Hill},
  \citenamefont {Kim}, \citenamefont {Hone}, \citenamefont {Li}, \citenamefont
  {Smirnov},\ and\ \citenamefont {Heinz}}]{Li2014}%
  \BibitemOpen
  \bibfield  {author} {\bibinfo {author} {\bibfnamefont {Y.}~\bibnamefont
  {Li}}, \bibinfo {author} {\bibfnamefont {J.}~\bibnamefont {Ludwig}}, \bibinfo
  {author} {\bibfnamefont {T.}~\bibnamefont {Low}}, \bibinfo {author}
  {\bibfnamefont {A.}~\bibnamefont {Chernikov}}, \bibinfo {author}
  {\bibfnamefont {X.}~\bibnamefont {Cui}}, \bibinfo {author} {\bibfnamefont
  {G.}~\bibnamefont {Arefe}}, \bibinfo {author} {\bibfnamefont {Y.~D.}\
  \bibnamefont {Kim}}, \bibinfo {author} {\bibfnamefont {A.~M.}\ \bibnamefont
  {van~der Zande}}, \bibinfo {author} {\bibfnamefont {A.}~\bibnamefont
  {Rigosi}}, \bibinfo {author} {\bibfnamefont {H.~M.}\ \bibnamefont {Hill}},
  \bibinfo {author} {\bibfnamefont {S.~H.}\ \bibnamefont {Kim}}, \bibinfo
  {author} {\bibfnamefont {J.}~\bibnamefont {Hone}}, \bibinfo {author}
  {\bibfnamefont {Z.}~\bibnamefont {Li}}, \bibinfo {author} {\bibfnamefont
  {D.}~\bibnamefont {Smirnov}}, \ and\ \bibinfo {author} {\bibfnamefont
  {T.~F.}\ \bibnamefont {Heinz}},\ }\href {\doibase
  10.1103/PhysRevLett.113.266804} {\bibfield  {journal} {\bibinfo  {journal}
  {Physical Review Letters}\ }\textbf {\bibinfo {volume} {113}},\ \bibinfo
  {pages} {266804} (\bibinfo {year} {2014})}\BibitemShut {NoStop}%
\bibitem [{\citenamefont {Srivastava}\ \emph {et~al.}(2015)\citenamefont
  {Srivastava}, \citenamefont {Sidler}, \citenamefont {Allain}, \citenamefont
  {Lembke}, \citenamefont {Kis},\ and\ \citenamefont
  {Imamoglu}}]{Srivastava2015}%
  \BibitemOpen
  \bibfield  {author} {\bibinfo {author} {\bibfnamefont {A.}~\bibnamefont
  {Srivastava}}, \bibinfo {author} {\bibfnamefont {M.}~\bibnamefont {Sidler}},
  \bibinfo {author} {\bibfnamefont {A.~V.}\ \bibnamefont {Allain}}, \bibinfo
  {author} {\bibfnamefont {D.~S.}\ \bibnamefont {Lembke}}, \bibinfo {author}
  {\bibfnamefont {A.}~\bibnamefont {Kis}}, \ and\ \bibinfo {author}
  {\bibfnamefont {A.}~\bibnamefont {Imamoglu}},\ }\href {\doibase
  10.1038/nphys3203} {\bibfield  {journal} {\bibinfo  {journal} {Nature
  Physics}\ }\textbf {\bibinfo {volume} {11}},\ \bibinfo {pages} {141}
  (\bibinfo {year} {2015})}\BibitemShut {NoStop}%
\bibitem [{\citenamefont {Aivazian}\ \emph {et~al.}(2015)\citenamefont
  {Aivazian}, \citenamefont {Gong}, \citenamefont {Jones}, \citenamefont {Chu},
  \citenamefont {Yan}, \citenamefont {Mandrus}, \citenamefont {Zhang},
  \citenamefont {Cobden}, \citenamefont {Yao},\ and\ \citenamefont
  {Xu}}]{Aivazian2015}%
  \BibitemOpen
  \bibfield  {author} {\bibinfo {author} {\bibfnamefont {G.}~\bibnamefont
  {Aivazian}}, \bibinfo {author} {\bibfnamefont {Z.}~\bibnamefont {Gong}},
  \bibinfo {author} {\bibfnamefont {A.~M.}\ \bibnamefont {Jones}}, \bibinfo
  {author} {\bibfnamefont {R.-L.~L.}\ \bibnamefont {Chu}}, \bibinfo {author}
  {\bibfnamefont {J.}~\bibnamefont {Yan}}, \bibinfo {author} {\bibfnamefont
  {D.~G.}\ \bibnamefont {Mandrus}}, \bibinfo {author} {\bibfnamefont
  {C.}~\bibnamefont {Zhang}}, \bibinfo {author} {\bibfnamefont
  {D.}~\bibnamefont {Cobden}}, \bibinfo {author} {\bibfnamefont
  {W.}~\bibnamefont {Yao}}, \ and\ \bibinfo {author} {\bibfnamefont {X.~D.}\
  \bibnamefont {Xu}},\ }\href {\doibase 10.1038/nphys3201} {\bibfield
  {journal} {\bibinfo  {journal} {Nature Physics}\ }\textbf {\bibinfo {volume}
  {11}},\ \bibinfo {pages} {148} (\bibinfo {year} {2015})}\BibitemShut
  {NoStop}%
\bibitem [{\citenamefont {MacNeill}\ \emph {et~al.}(2015)\citenamefont
  {MacNeill}, \citenamefont {Heikes}, \citenamefont {Mak}, \citenamefont
  {Anderson}, \citenamefont {Korm{\'{a}}nyos}, \citenamefont {Z{\'{o}}lyomi},
  \citenamefont {Park},\ and\ \citenamefont {Ralph}}]{MacNeill2015}%
  \BibitemOpen
  \bibfield  {author} {\bibinfo {author} {\bibfnamefont {D.}~\bibnamefont
  {MacNeill}}, \bibinfo {author} {\bibfnamefont {C.}~\bibnamefont {Heikes}},
  \bibinfo {author} {\bibfnamefont {K.~F.}\ \bibnamefont {Mak}}, \bibinfo
  {author} {\bibfnamefont {Z.}~\bibnamefont {Anderson}}, \bibinfo {author}
  {\bibfnamefont {A.}~\bibnamefont {Korm{\'{a}}nyos}}, \bibinfo {author}
  {\bibfnamefont {V.}~\bibnamefont {Z{\'{o}}lyomi}}, \bibinfo {author}
  {\bibfnamefont {J.}~\bibnamefont {Park}}, \ and\ \bibinfo {author}
  {\bibfnamefont {D.~C.}\ \bibnamefont {Ralph}},\ }\href {\doibase
  10.1103/PhysRevLett.114.037401} {\bibfield  {journal} {\bibinfo  {journal}
  {Physical Review Letters}\ }\textbf {\bibinfo {volume} {114}},\ \bibinfo
  {pages} {037401} (\bibinfo {year} {2015})}\BibitemShut {NoStop}%
\bibitem [{\citenamefont {Popescu}\ and\ \citenamefont
  {Zunger}(2010)}]{Popescu2010}%
  \BibitemOpen
  \bibfield  {author} {\bibinfo {author} {\bibfnamefont {V.}~\bibnamefont
  {Popescu}}\ and\ \bibinfo {author} {\bibfnamefont {A.}~\bibnamefont
  {Zunger}},\ }\href {\doibase 10.1103/PhysRevLett.104.236403} {\bibfield
  {journal} {\bibinfo  {journal} {Physical Review Letters}\ }\textbf {\bibinfo
  {volume} {104}},\ \bibinfo {pages} {236403} (\bibinfo {year}
  {2010})}\BibitemShut {NoStop}%
\bibitem [{\citenamefont {Wigner}(1934)}]{Wigner1934}%
  \BibitemOpen
  \bibfield  {author} {\bibinfo {author} {\bibfnamefont {E.}~\bibnamefont
  {Wigner}},\ }\href {\doibase 10.1103/PhysRev.46.1002} {\bibfield  {journal}
  {\bibinfo  {journal} {Physical Review}\ }\textbf {\bibinfo {volume} {46}},\
  \bibinfo {pages} {1002} (\bibinfo {year} {1934})}\BibitemShut {NoStop}%
\bibitem [{\citenamefont {Grimes}\ and\ \citenamefont
  {Adams}(1979)}]{Grimes1979}%
  \BibitemOpen
  \bibfield  {author} {\bibinfo {author} {\bibfnamefont {C.~C.}\ \bibnamefont
  {Grimes}}\ and\ \bibinfo {author} {\bibfnamefont {G.}~\bibnamefont {Adams}},\
  }\href {\doibase 10.1103/PhysRevLett.42.795} {\bibfield  {journal} {\bibinfo
  {journal} {Physical Review Letters}\ }\textbf {\bibinfo {volume} {42}},\
  \bibinfo {pages} {795} (\bibinfo {year} {1979})}\BibitemShut {NoStop}%
\bibitem [{\citenamefont {Cooper}(1996)}]{Cooper1996}%
  \BibitemOpen
  \bibfield  {author} {\bibinfo {author} {\bibfnamefont {N.~R.}\ \bibnamefont
  {Cooper}},\ }\href {\doibase 10.1103/PhysRevB.53.10804} {\bibfield  {journal}
  {\bibinfo  {journal} {Physical Review B}\ }\textbf {\bibinfo {volume} {53}},\
  \bibinfo {pages} {10804} (\bibinfo {year} {1996})}\BibitemShut {NoStop}%
\bibitem [{\citenamefont {Kn{\"{o}}rzer}\ \emph {et~al.}(2020)\citenamefont
  {Kn{\"{o}}rzer}, \citenamefont {Schuetz}, \citenamefont {Giedke},
  \citenamefont {Wild}, \citenamefont {{De Greve}}, \citenamefont {Schmidt},
  \citenamefont {Lukin},\ and\ \citenamefont {Cirac}}]{Knorzer2020}%
  \BibitemOpen
  \bibfield  {author} {\bibinfo {author} {\bibfnamefont {J.}~\bibnamefont
  {Kn{\"{o}}rzer}}, \bibinfo {author} {\bibfnamefont {M.~J.~A.}\ \bibnamefont
  {Schuetz}}, \bibinfo {author} {\bibfnamefont {G.}~\bibnamefont {Giedke}},
  \bibinfo {author} {\bibfnamefont {D.~S.}\ \bibnamefont {Wild}}, \bibinfo
  {author} {\bibfnamefont {K.}~\bibnamefont {{De Greve}}}, \bibinfo {author}
  {\bibfnamefont {R.}~\bibnamefont {Schmidt}}, \bibinfo {author} {\bibfnamefont
  {M.~D.}\ \bibnamefont {Lukin}}, \ and\ \bibinfo {author} {\bibfnamefont
  {J.~I.}\ \bibnamefont {Cirac}},\ }\href {\doibase
  10.1103/PhysRevB.101.125101} {\bibfield  {journal} {\bibinfo  {journal}
  {Physical Review B}\ }\textbf {\bibinfo {volume} {101}},\ \bibinfo {pages}
  {125101} (\bibinfo {year} {2020})}\BibitemShut {NoStop}%
\bibitem [{\citenamefont {Koulakov}\ \emph {et~al.}(1996)\citenamefont
  {Koulakov}, \citenamefont {Fogler},\ and\ \citenamefont
  {Shklovskii}}]{Koulakov1996}%
  \BibitemOpen
  \bibfield  {author} {\bibinfo {author} {\bibfnamefont {A.~A.}\ \bibnamefont
  {Koulakov}}, \bibinfo {author} {\bibfnamefont {M.~M.}\ \bibnamefont
  {Fogler}}, \ and\ \bibinfo {author} {\bibfnamefont {B.~I.}\ \bibnamefont
  {Shklovskii}},\ }\href {\doibase 10.1103/PhysRevLett.76.499} {\bibfield
  {journal} {\bibinfo  {journal} {Physical Review Letters}\ }\textbf {\bibinfo
  {volume} {76}},\ \bibinfo {pages} {499} (\bibinfo {year} {1996})}\BibitemShut
  {NoStop}%
\bibitem [{\citenamefont {Moessner}\ and\ \citenamefont
  {Chalker}(1996)}]{Moessner1996}%
  \BibitemOpen
  \bibfield  {author} {\bibinfo {author} {\bibfnamefont {R.}~\bibnamefont
  {Moessner}}\ and\ \bibinfo {author} {\bibfnamefont {J.~T.}\ \bibnamefont
  {Chalker}},\ }\href {\doibase 10.1103/PhysRevB.54.5006} {\bibfield  {journal}
  {\bibinfo  {journal} {Physical Review B}\ }\textbf {\bibinfo {volume} {54}},\
  \bibinfo {pages} {5006} (\bibinfo {year} {1996})}\BibitemShut {NoStop}%
\end{thebibliography}
\end{document}


\begin{center}
{\Large Supplemental Material}\\
\end{center}
\tableofcontents

\newpage

\section{E\lowercase{xtraction of energy separation between exciton and \uppercase{U}mklapp exciton resonances}}
The magnitude of the oscillator strength of Umklapp exciton is small and  the extraction of its resonance energy from differential reflectance ($\Delta R/R_0$) measurements is difficult. This difficulty arises from the dispersive $\Delta R/R_0$ lineshape, which in turn is a consequence of interference between the field generated by resonantly driven excitons and reflection of the drive field from various interfaces in the sample. The latter contribution to $\Delta R/R_0$ has the same polarization as the incident light whereas the field generated by the exciton is linearly polarized for samples with uniaxial strain. 

To overcome the limitation that is inherent to $\Delta R/R_0$ measurements, we have carried out resonance fluorescence (RF) measurements to accurately determine the energy separation between main exciton resonance and the Umklapp exciton peak stemming from the periodic potential due to electrons in a Mott state. RF measurements faithfully reproduce the Lorentzian lineshape of the excitonic resonances with a definite polarization axis. To implement RF measurements, the sample is excited with linearly polarized white light source (broad band light-emitting diode with a center wavelength of 760~nm, and a band width of 20~nm), and the generated cross-polarized light is detected by filtering out the reflected drive field using a polarizer. Since light reflection at dielectric interfaces at or near normal incidence preserves the polarization, the photons detected in this configuration stem predominantly from excitons. To the extent that the polarization degeneracy of the excitonic resonance is lifted either by uniaxial strain or by an external magnetic field, it is possible to detect up to $50~\%$ of the photons generated by exciton radiative decay. Since the RF lineshape is Lorentzian with tails that decay quadratically with detuning from the center exciton frequency, the identification of the Umklapp peaks become much easier. In our experiments, we selected the polarization angle to ensure that we have signal from both top ($\rm X_{top}$) and bottom ($\rm X_{bot}$) layer exciton resonances in the charge neutral regime, and normalized the signal by the weight of the excitation white light intensity distribution.

Figure S1(a) shows the comparison of RF signal, $\Delta R/R_0$ and the derivative of $\Delta R/R_0$ with respect to emission energy ($d(\Delta R/R_0)/dE$) at $\nu = 1$, where $(\nu_{\rm top}, \nu_{\rm bot}) = (1, 0)$.
$\rm X_{top}$ and $\rm X_{bot}$ exciton resonances in RF data correspond to the resonances in the reflectance signal.
The top layer Umklapp exciton peak resonance is identified as a small peak indicated with $\rm X_{top}^U$ in RF measurement (top panel). The same resonance shows a kink in the $\Delta R/R_0$ and a minima in $d(\Delta R/R_0)/dE$ (bottom panel). The energy splitting between $\rm X_{top}$ and $\rm X_{top}^U$ obtained from the local maxima of these RF resonances is 2.7~meV.
Figure S1(b) shows the RF signal and $\Delta R/R_0$ signal at $\nu = 1$, where $(\nu_{\rm top}, \nu_{\rm bot}) = (0, 1)$.

The Umklapp exciton resonance from the bottom layer is indicated by $\rm X_{bot}^U$ (top panel). The same resonance is also visible as a kink in $\Delta R/R_0$ and a minima in $d(\Delta R/R_0)/dE$ (bottom panel).
The energy splitting between $\rm X_{bot}$ and $\rm X_{bot}^U$ estimated from RF data is 2.7~meV.

Figure S1(c) shows the RF (top panel) and $\Delta R/R_0$ (bottom panel) measurements at $\nu = 2$, where $(\nu_{\rm top}, \nu_{\rm bot}) = (1, 1)$.
Similarly to the above cases, the Umklapp exciton resonances in RF data are indicated with $\rm X_{top}^U$ and $\rm X_{bot}^U$, respectively. At $(\nu_{\rm top}, \nu_{\rm bot}) = (1, 1)$, the energy splitting between $\rm X_{top}$ and $\rm X_{top}^U$ is 2.8~meV and the energy splitting between $\rm X_{bot}$ and $\rm X_{bot}^U$ is 2.6~meV. 

The RF measurements depicted in Fig.~S1 clearly shows the power of RF measurements in accurately identifying the energy splitting between the main and Umklapp exciton peaks. However, these measurements also reveal the presence of several other weak resonances that are not discernible in $\Delta R/R_0$ measurements. The identification of these weaker peaks will be the subject of future work.

\newpage

\begin{figure*}[h]
\centering
\includegraphics[width=1.0\textwidth]{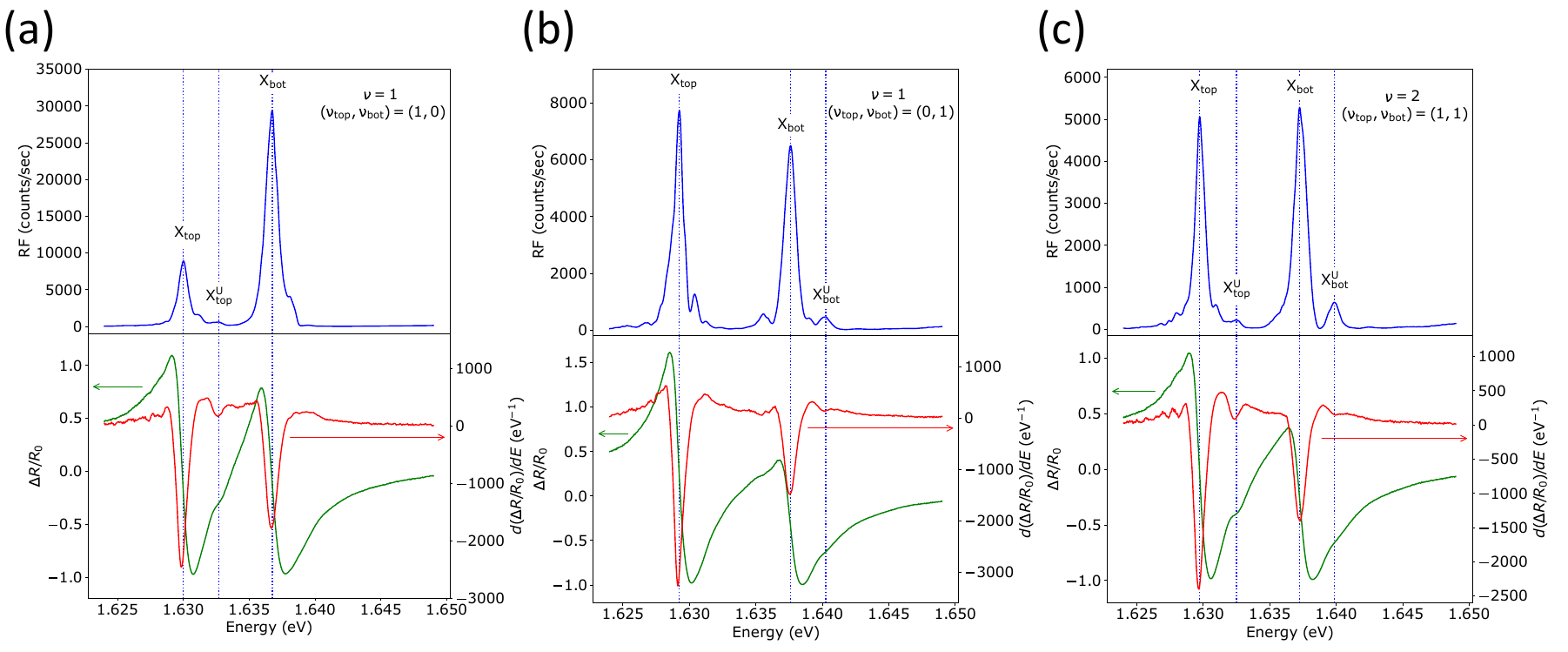}
\caption{
Comparison between resonance fluorescence and differential reflectance.
(a) - (c) Top panels show the RF data and the bottom panels show the differential reflectance ($\Delta R/R_0$) and energy derivative of $\Delta R/R_0$, for the cases of $(\nu_{\rm top}, \nu_{\rm bot}) = (1, 0)$ (a), $(\nu_{\rm top}, \nu_{\rm bot}) = (0, 1)$ (b), and $(\nu_{\rm top}, \nu_{\rm bot}) = (1, 1)$ (c).
The blue dashed lines indicate the energies where the RF signals are locally maximum around the indicated resonances.
All RF, $\Delta R/R_0$, and $d(\Delta R/R_0)/dE$ signals are averaged in the range of $V_{\rm E} = 0.166 \sim 0.314~{\rm V}$ for $(\nu_{\rm top}, \nu_{\rm bot}) = (1, 0)$, $V_{\rm E} = -2.054 \sim -1.906~{\rm V}$ for $(\nu_{\rm top}, \nu_{\rm bot}) = (0, 1)$, and , $V_{\rm E} = -0.870 \sim -0.722~{\rm V}$ for $(\nu_{\rm top}, \nu_{\rm bot}) = (1, 1)$.
RF and $d(\Delta R/R_0)/dE$ data are smoothed taking moving average with the energy window of 0.16~meV and 0.24~meV, respectively.
}
\label{figS1}
\end{figure*}

\newpage

\section{E\lowercase{xtraction of electron filling periodicity}}
Electron filling periodicity is extracted from the periodic energy shift of exciton/repulsive polaron resonances along the $V_{\rm \mu}$ axis.
For the extraction of these resonance energies from the differential reflectance spectrum ($\Delta R/R_0$), we performed the following analysis.
Each exciton/repulsive polaron resonance ($\rm X_{L}$, L = top, bot) is modeled with the following phenomenological dispersive Lorentzian formula\cite{Smolenski2019},
\begin{equation}
    L(E; E_{\rm X_L}, \gamma_{\rm X_L}, A_{\rm X_L}, \alpha_{\rm X_L}) =
    A\cos{\alpha_{\rm X_L}}\frac{\gamma_{\rm X_L}/2}{(E-E_{\rm X_L})^2+\gamma_{\rm X_L}^2/4} +
    A\sin{\alpha_{\rm X_L}}\frac{E_{\rm X_L} - E}{(E-E_{\rm X_L})^2+\gamma_{\rm X_L}^2/4},
\end{equation}
where $E_{\rm X_L}$, $\gamma_{\rm X_L}$, $A_{\rm X_L}$, $\alpha_{\rm X_L}$ represent the resonance energy, linewidth, amplitude, and phase shift of the resonance, and $E$ is the energy as a variable of the function.
To extract $E_{\rm X_{top}}$ and $E_{\rm X_{bot}}$, the $\Delta R/R_0$ spectra are fitted with
\begin{equation}
    \frac{\Delta R}{R_0}(E) = L(E; E_{\rm X_{top}}, \gamma_{\rm X_{top}}, A_{\rm X_{top}}, \alpha_{\rm X_{top}}) + L(E; E_{\rm X_{bot}}, \gamma_{\rm X_{bot}}, A_{\rm X_{bot}}, \alpha_{\rm X_{bot}}) + C,
\end{equation}
where $C$ is the constant to describe the offset reflectance.
We note that the Umklapp exciton play a negligible for the extraction of the main exciton/repulsive polaron resonance energies, due to their small oscillator strength.
Figure S2 shows the extracted $E_{\rm X_{bot}}$ differentiated with respect to $V_{\rm \mu}$.
While the bottom layer is filled, the corresponding repulsive polaron resonance exhibits a blue shift ($dE_{\rm X_{bot}}/dV_{\rm \mu}>0$) and we can identify the checker board pattern emerging from layer by layer filling of electrons.
From the periodicity, we assign the fillings $\nu = 1, 2, 3, 4$ which are
indicated by the blue dashed lines in Fig.~S2.
The corresponding density periodicity is $n_{\rm moir\acute{e}} \sim 1.9 \times 10^{11}~{\rm cm^{-2}}$ with about $15~\%$ uncertainty, which matches the value we have estimated in our previous study based on the same sample~\cite{Shimazaki2020}.
The \moire\ periodicity $a_{\rm M}$ is given by the relation $(\sqrt{3}/2) a_{\rm M}^2 = 1/n_{\rm moir\acute{e}}$, which results in $a_{\rm M} = 25 \pm 2 {\rm nm}$.
Please note that we change the definition of $\nu$ from the previous work (Ref. \cite{Shimazaki2020}) to be consistent with the definition by most of the studies of twisted bilayer graphene. Here $\nu = 1$ is one electron per \moire\ site, which is half filling of the \moire\ band.

\begin{figure*}[b]
\centering
\includegraphics[width=0.4\textwidth]{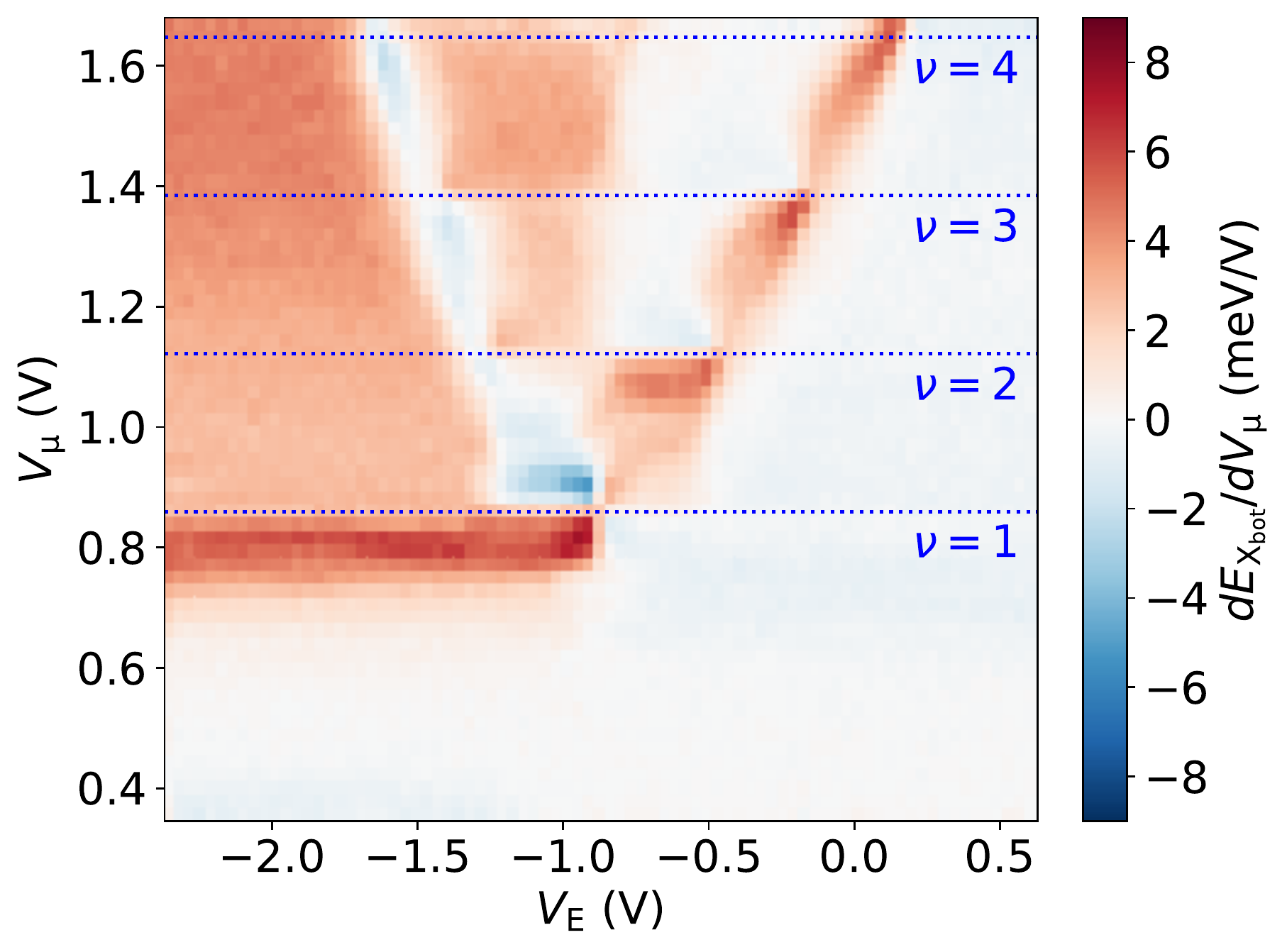}
\caption{
Bottom layer exciton/repulsive polaron resonance energy differentiated with respect to $V_{\rm \mu}$.
$V_{\rm E}$ and $V_{\rm \mu}$ dependence of the bottom layer exciton/repulsive polaron resonance energy ($E_{\rm X_{bot}}$) differentiated with respect to $V_{\rm \mu}$.
}
\label{figS2}
\end{figure*}

\newpage
\section{B\lowercase{lue shift of the main repulsive polaron resonance in a \uppercase{M}ott state}}
For $V_{\rm E}$ values corresponding to $E_z \simeq  0$, the appearance of a Mott state in a given layer is also associated with a cusp-like blue shift in the main repulsive polaron resonance~\cite{Smolenski2019}: this is particularly striking in Fig.~3(a) in the main text where the cusp in top layer exciton at $\nu = 1$ is accompanied with a weak red-shift of the bottom layer exciton peak. The transfer of electrons from the bottom layer to the top layer as $\nu$ is increased from $0.7$ to $1.0$ is at least partly responsible for these energy shifts. However, the sharp blue-shift of the top layer exciton in the immediate neighbor hood of $\nu=1$, as well as the origin of the weak lower energy resonance most likely requires a theoretical analysis that captures dynamical screening of excitons by elementary collective excitations of electrons in a Mott state.

\section{F\lowercase{illing dependence of \uppercase{U}mklapp exciton resonances}}
In the main text, we discuss the appearance of Umklapp exciton resonances at $\nu = 1$ and 2.
Here we show additional data set characterizing the $V_{\rm E}$ dependence of $\Delta R/R_0$ for each fixed $\nu$.
In Fig.~S3, we show the filling dependence of $\Delta R/R_0$ spectrum as well as of $d(\Delta R/R_0)/dE$ around $\nu = 1$.
Comparing Fig.~S3, (a) to (c), we can clearly find that the shift of exciton resonances is sharp at $\nu = 1$, and smooth out for higher and lower fillings, which is the signature of formation of Mott state at $\nu = 1$ (see also Ref.  \cite{Shimazaki2020}).
As we discuss in the main text, at $\nu = 1$ [Fig.~S3(b)], upon the transition between $(\nu_{\rm top}, \nu_{\rm bot}) = (1, 0)$ and $(0, 1)$ around $V_{\rm E} \sim -0.9~{\rm V}$, there are corresponding Umklapp exciton resonances at this filling [Fig.~S3(e)].
In lower filling [$\nu = 0.67$, Fig.~S3(d)] and higher filling [$\nu = 1.33$, Fig.~S3(f)], we observe that the Umklapp exciton resonances disappear.
Figure S4 shows the corresponding filling dependence of Umklapp exciton resonances around $\nu = 2$. We confirm that Umklapp exciton resonances at $\nu = 2$ [Fig.~S4(e)] in the $(\nu_{\rm top}, \nu_{\rm bot}) = (1, 1)$ region become less visible for lower filling [$\nu = 1.67$, Fig.~S4(d)] and higher filling [$\nu = 2.33$, Fig.~S4(f)].
These data demonstrate that the emergence of incompressible electronic crystal results in Umklapp exciton resonances.

In Fig.~S5, we show the $\Delta R/R_0$ and $d(\Delta R/R_0)/dE$ data for the filling factor $\nu \geq 3$. In this case, Umklapp resonances exist for both for $(\nu_{\rm top}, \nu_{\rm bot}) = (2, 1)$ and $(1, 2)$, but $\rm X_{top} - X_{top}^U$ (and also $\rm X_{bot} - X_{bot}^U$) energy separation is smaller when we have two electrons in the corresponding layer [Fig.~S5(d)]. This observation indicates that a different type of potential is realized for excitons, and when we have two electrons in one layer and one electron in the other layer per \moire\ site, one possible arrangement is that the layer which has two electrons forms hexagonal lattice and the other layer forms triangular lattice to avoid each other and create a triangular lattice in total.
For the case of $\nu = 4$, we also observe Umklapp resonances for both cases of $(\nu_{\rm top}, \nu_{\rm bot}) = (3, 1)$ and $(1, 3)$, but the underlying electronic structure is unclear [Fig.~S5(f)].
In between $\nu = 3$ and $\nu = 4$, we find that the Umklapp exciton resonances remain visible and correspond to those of $\nu = 3$ and $\nu = 4$, suggesting that the corresponding layer remains in an incompressible state [Fig.~S5(e)].

\newpage


\newpage

\newpage

\begin{figure*}[h]
\centering
\includegraphics[width=1.0\textwidth]{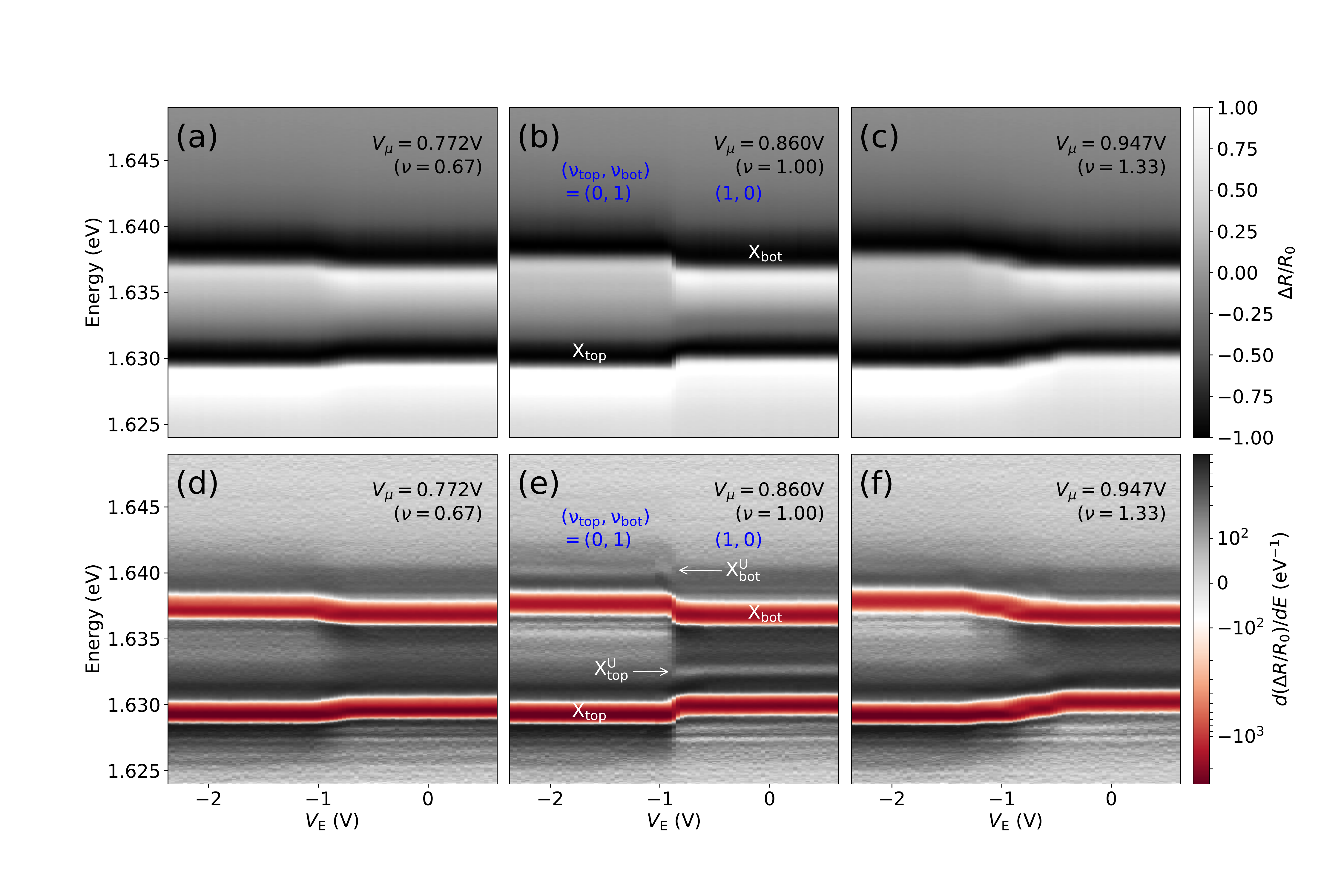}
\caption{
Filling dependence of Umklapp exciton resonances around $\nu = 1$.
(a) - (c) $V_{\rm E}$ dependence of the differnetial reflectance spectrum at $\nu = 0.67$ (a), 1.00 (b), and 1.33 (c).
(d) - (f) $V_{\rm E}$ dependence of the differnetial reflectance spectrum differentiated with respect to energy $E$ at $\nu = 0.67$ (d), 1.00 (e), and 1.33 (f). The scale of the color bar is logarithmic for $|d(\Delta R/R_0)/dE| > 10^2~{\rm eV^{-1}}$ and linear for $|d(\Delta R/R_0)/dE| < 10^2~{\rm eV^{-1}}$.
}
\label{figS3}
\end{figure*}

\newpage

\begin{figure*}[h]
\centering
\includegraphics[width=1.0\textwidth]{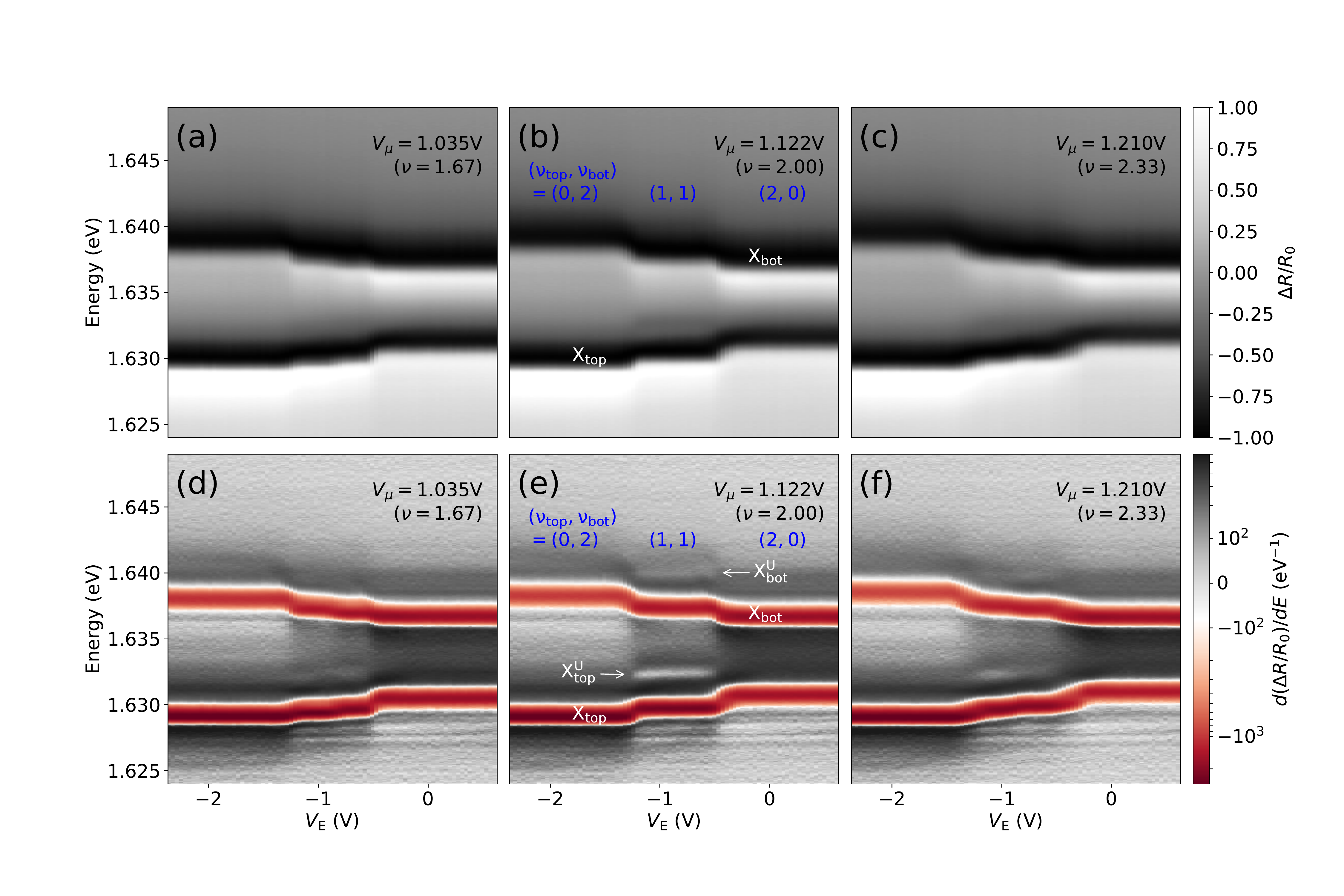}
\caption{
Filling dependence of Umklapp exciton resonances around $\nu = 2$.
(a) - (c) $V_{\rm E}$ dependence of the differnetial reflectance spectrum at $\nu = 1.67$ (a), 2.00 (b), and 2.33 (c).
(d) - (f) $V_{\rm E}$ dependence of the differnetial reflectance spectrum differentiated with respect to energy $E$ at $\nu = 1.67$ (d), 2.00 (e), and 2.33 (f). The scale of the color bar is logarithmic for $|d(\Delta R/R_0)/dE| > 10^2~{\rm eV^{-1}}$ and linear for $|d(\Delta R/R_0)/dE| < 10^2~{\rm eV^{-1}}$.
}
\label{figS4}
\end{figure*}

\newpage

\begin{figure*}[h]
\centering
\includegraphics[width=1.0\textwidth]{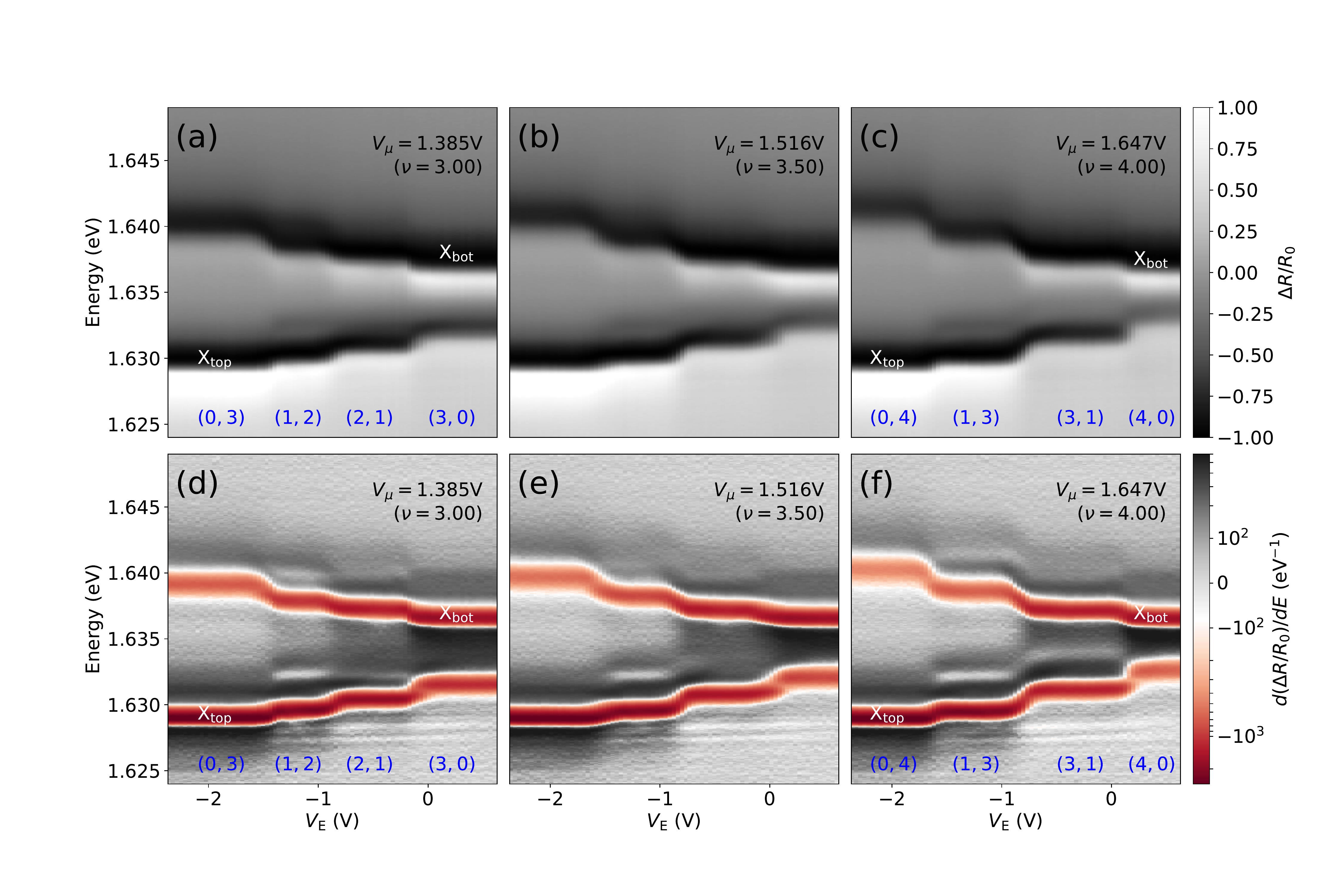}
\caption{
Filling dependence of Umklapp exciton resonances in higher filling $\nu \geq 3$.
(a) - (c) $V_{\rm E}$ dependence of the differnetial reflectance spectrum at $\nu = 3.00$ (a), 3.50 (b), and 4.00 (c).
(d) - (f) $V_{\rm E}$ dependence of the differnetial reflectance spectrum differentiated with respect to energy $E$ at $\nu = 3.00$ (d), 3.50 (e), and 4.00 (f). The scale of the color bar is logarithmic for $|d(\Delta R/R_0)/dE| > 10^2~{\rm eV^{-1}}$ and linear for $|d(\Delta R/R_0)/dE| < 10^2~{\rm eV^{-1}}$.
}
\label{figS5}
\end{figure*}

\newpage


%